\documentclass[num-refs]{wiley-article}

\bibstyle{WileyNJD-ACS}
\makeatletter
\renewcommand{\@biblabel}[1]{#1.}
\makeatother

\usepackage{siunitx}

\usepackage{mathrsfs}
\usepackage{soul}
\usepackage{cancel}
\usepackage[normalem]{ulem}

\papertype{Original Article}
\paperfield{Journal Section}

\title{Quantum density matrix theory for a laser without adiabatic elimination of the population inversion: transition to lasing in the class-B limit}

\author[1]{Alejandro M. Yacomotti}
\author[2]{Zakari Denis}
\author[3]{Alberto Biella}
\author[2]{Cristiano Ciuti}

\affil[1]{Centre de Nanosciences et de Nanotechnologies, CNRS, Universit\'e Paris-Saclay, 10 Boulevard Thomas Gobert, 91120 Palaiseau, France}
\affil[2]{Universit\'{e} Paris Cit\'{e}, CNRS, Laboratoire Mat\'{e}riaux et Ph\'{e}nom\`{e}nes Quantiques (MPQ), F-75013 Paris, France}
\affil[3]{INO-CNR BEC Center and Dipartimento di Fisica, Universit\`a di Trento, 38123 Povo, Italy}

\corraddress{Alejandro M. Yacomotti}
\corremail{alejandro.giacomotti@c2n.upsaclay.fr}

\fundinginfo{This work was supported by ANR via the projects UNIQ (ANR-16-CE24-0029), NOMOS (ANR-18-CE24-0026), TRIANGLE (ANR-20-CE47-0011), the FET FLAGSHIP Project PhoQuS (grant agreement ID: 820392) and  the European Union in the form of Marie Sk\l odowska-Curie Action grant MSCA-841351.}

\runningauthor{A. M. Yacomotti et al.}

\begin{document}

\begin{frontmatter}
\maketitle

\begin{abstract}
Despite the enormous technological interest in micro and nanolasers, surprisingly, no class-B quantum density-matrix model is available to date, capable of accurately describing coherence and photon correlations within a unified theory.
In class-B lasers ---applicable for most solid--state lasers at room temperature---, the macroscopic polarization decay rate is larger than the cavity damping rate which, in turn, exceeds the upper level population decay rate. 
Here we carry out a density-matrix theoretical approach for generic class-B lasers, and provide closed equations for the photonic and atomic reduced density matrix in the Fock basis of photons.
Such a relatively simple model can be numerically integrated in a straightforward way, and exhibits all the expected phenomena, from one-atom photon antibunching, to the well-known S-shaped input-output laser emission and super-Poissonian autocorrelation for many atoms ($1\leq g^{(2)}(0)\leq 2$), and from few photons (large spontaneous emission factors, $\beta\sim1$) to the thermodynamic limit ($N\gg1$ and $\beta\sim 0$). Based on the analysis of $g^{(2)}(\tau)$, we conclude that super-Poissonian fluctuations are clearly related to relaxation oscillations in the photon number. We predict a strong damping of relaxation oscillations
with an atom number as small as $N\sim 10$. This model enables the study of few-photon bifurcations and non-classical photon correlations in class-B laser devices, also leveraging quantum descriptions of coherently coupled nanolaser arrays.

\end{abstract}
\end{frontmatter}

\section{Introduction}
Quantum density-matrix theory gives a handle to access the full statistical information of open manybody quantum systems. In laser physics, a density-matrix approach for class-A lasers has been derived in the 80s by M. Scully and co-workers \cite{scully1997quantum}. Class-A lasers include He-Ne, Ar, dye lasers, but also nanolasers with ultrahigh Q-factors \cite{Takemura_2021}. In laser-physics terminology, class-A means that the cavity damping rate $\kappa$ is much smaller than the relaxation rates of the atomic variables, $\kappa\ll \gamma_\parallel, \gamma_\perp$, where $\gamma_\parallel$ is the spontaneous decay rate of the laser's upper state, and $\gamma_\perp$ the damping rate of the material's macroscopic polarization \cite{arecchi2012instabilities}. Therefore, the atomic variables can be adiabatically eliminated, and cavity-induced dissipation, as well as atom-atom correlations, can be neglected, strongly simplifying the dynamical equations and enabling closed-form analytical solutions of the Linblad master equation. Importantly, the density-matrix approach provides access to the full photon statistics and coherence, which are hindered in most standard theoretical approaches such as semiclassical Maxwell-Bloch equations, rate equations or birth-death models. 

Despite the simplicity and elegance of class-A quantum laser theory, the great majority of solid-state devices operating at room temperature, such as semiconductor lasers, including most micro and nanolasers, belong to class-B. Within this class, $\gamma_\perp\gg \kappa > \gamma_\parallel$, therefore the population inversion variable can no longer be adiabatically eliminated. 
Importantly, there has been a growing interest in the past three decades in micro and nanoscale semiconductor lasers for integrated photonics, both for fundamentals and applications. It is well known that spontaneous emission effects, quantified through the spontaneous emission factor $\beta$, are dominant in small cavities because of the large amount of spontaneous emission coupled into the laser mode, with dramatic consequences in the spectral and photon-statistical observables (thresholdless regime). 

A large amount of theoretical models have been developed in the past twenty years to describe the transition to lasing and the building-up of coherence in optical cavities with gain media, including semiconductor microcavities, with particular emphasis in the so-called cQED regime (strong coupling). Those approaches range from semiclassical Maxwell-Bloch and rate equations \cite{narducci1988laser,PhysRevA.50.4318}, Heisenberg-Langevin equations \cite{PhysRevA.44.657,PhysRevA.59.1667}, birth-death models \cite{PhysRevA.50.4318,PhysRevA.87.053819}, Jaynes-Cummings models with discrete emitters \cite{6603264} to semiconductor microscopic theory  \cite{PhysRevA.75.013803,Wiersig_2009,KIRA1999189,chow2014emission,2017LSA.....6E7030K,PhysRevA.87.053819,PhysRevB.87.205310}. In the latter, populations are determined by using Heisenberg's equations of motion, generally solved within a cluster-expansion scheme; such a theory has been particularly useful for describing semiconductor microcavities in the cQED regime \cite{Wiersig_2009} since it is able to incorporate specific features of semiconductors such as Pauli blocking and Coulomb interactions which, in turn, make those models very complex. Recently, this kind of approach has been applied for describing laser transition and photon correlations within a bifurcation analysis for an ensemble of identical emitters ---from a few to a very large number--- leading to the prediction of collective antibunching before coherent laser emission \cite{PhysRevLett.126.063902}, which had also been reported in cavities with few QDs \cite{chow2014emission}. 
In spite of the substantial efforts and progress for modeling correlations in lasers in the recent years ---some of them particularly successful---, an unifying understanding of the laser transition of devices, also valid at the forefront of technology, is still elusive. A density-matrix theory can be considered as an universal approach since it bridges the gap between laser physics and the more general theory of dissipative phase transitions  \cite{Minganti_2018,Minganti_2021,Takemura_2021}. 

In this work we extend Scully's density-matrix approach into the class-B regime, and provide closed-form equations of motion for the photonic and atomic reduced density matrix in the Fock basis of photons. We show that in the class-B regime, two main ingredients can no longer be neglected: (i) the cavity-induced dissipation terms in the atomic variables, and (ii) the two-atom contributions in the gain terms, both missing in class-A models. Because of these two effects, analytical solutions as in class-A lasers are no longer available, but the resulting dynamical system for the two coupled reduced density-matrix elements ---the atomic and photonic ones--- are relatively simple and can be numerically solved with standard methods. 
In particular, we achieve this goal by introducing a conditional-probability ansatz for atom-atom correlations that enables a closure of the system of equations. We stress that this \lq\lq atomic-like'' model can be applied to several gain materials with large pure dephasing, not only semiconductors but also ion-doped crystals and glasses (optical fibers) operating at room temperature.   

Our model accurately describes a plethora of light-emission phenomena, starting from photon antibunching in the one-atom limit ($g^{(2)}(0)<1$). On the opposite ---thermodynamic--- limit (large atomic populations and small $\beta$-factors), the zero time delay second-order coherence $g^{(2)}(0)$ sharply and monotonically decreases from  $g^{(2)}(0)\approx 2$ to $g^{(2)}(0)\approx 1$ when crossing the laser threshold, which is well-described with standard models. Another result well-captured by this model is the slow evolution of the photon-number fluctuations towards coherence [$g^{(2)}(0)=1$] in the $\beta \sim 1$ regime, in agreement with previous analysis in terms of rate equations and birth-death models \cite{PhysRevA.50.4318}. The time-lag-dependent autocorrelation $g^{(2)}(\tau)$ can be readily computed, and confirms that super-Poissonian light in the $\beta \sim 1$ regime is related to intensity noise excess from relaxation oscillations in the photon number, as it has already been suggested for class-B nanolasers \cite{PhysRevLett.110.163603}. Somehow unexpectedly, the model predicts that the emission of class-B nanolasers with few-intracavity atoms ($N\sim 10$) remains super-Poissonian, while the relaxation oscillations become strongly damped compared to large $N$ devices. 

This paper is organized in three main sections. In Sec. \ref{Class-A lasers}, we review the quantum density-matrix approach for class-A lasers developed in \cite{scully1997quantum}, with special emphasis in highlighting the role of the number of atoms $N$, which had therein been absorbed in the normalization choice; we discuss the role of $N$ in the gain terms and in the laser threshold. The original contribution of this work is presented in Sec. \ref{Class-B lasers}, where the dynamical equations for class-B lasers are derived. Numerical results are shown and systematically compared with "class-A-like" solutions, and reveal that the main difference lies in the photon-number fluctuations, quantified through the second-order coherence [$g^2(0)$]. Two limit cases of laser operation will be discussed: the single-atom cavity and the thermodynamic limit (large $N$ and small $\beta$). Although the former cannot operate above threshold within class-B, it displays strong photon antibunching as expected, showing that this model is able to capture non-classical fluctuations of light. In addition, two micro and nanolaser examples will be computed and discussed: the mesoscopic regime with $\beta\sim 0.01$ and $N=10^5$, and the nanoscopic thresholdless regime ($N=10$ and $\beta\sim 1$). Finally, we compute the dynamics of second-order correlations [$g^{(2)}(\tau)$] and systematically compare class-B with class-A-like model predictions in Sec. \ref{Correlations}. Relaxation oscillations are clearly present in the class-B model, while they are absent in the class-A-like model, as expected. We conclude our work and discuss some prospects in Sec. \ref{Conclusions}.

\section{Class-A lasers}
\label{Class-A lasers}

In this section, we review the results for a class-A laser system derived by Scully and co-workers \cite{scully1997quantum}. We consider a system of $N$ atoms inside an optical cavity. We will explicitly discuss the role of $N$ and the self-consistency of the approach. The model in Ref. \cite{scully1997quantum} accounts for the general case of a five-level atomic system, which is eventually reduced to the three-level system represented in Fig. 1: the laser transition takes place between the upper level $|a\rangle$ and the lower level $|b\rangle$, while $|g\rangle$ is the ground state. The population in level $|a\rangle$ ($|b\rangle$) decays to the state $|g\rangle$ with a rate $\gamma_{a(b)}$. 

\subsection{Density matrix approach for a class-A laser}

Under the rotating-wave approximation, the light-matter interaction can be modeled by the following Hamiltonian:
\begin{equation}
\hat{\mathscr{V}}=\sum_{i=1}^{N} \hbar g \left( \hat \sigma_+^i \hat a+\hat a^\dagger \hat \sigma_- ^i \right) =\sum_{i=1}^{N} \hat{\mathscr{V}}_i,
\label{eq:Hamiltonian}
\end{equation}
where $\hat \sigma_+^i=(|a\rangle \langle b|)^i$ and $\hat \sigma_-^i=(|b\rangle \langle a|)^i$ are the rising and lowering operators for the $i$th atom for the $|a\rangle \leftrightarrow  |b\rangle$ transition. The equation of motion of the density matrix in the presence of cavity and atomic damping mechanisms is 
\begin{equation}
\dot {\hat \rho} =-\frac{i}{\hbar} [\hat{\mathscr{V}},\hat \rho]+{\cal L}_\mathrm{cav}\hat \rho+{\cal L}_\mathrm{atom} \hat \rho,
\label{eq:master}
\end{equation}
where ${\cal L}_\mathrm{cav}$ and ${\cal L}_\mathrm{atom}$ are the Linblad superoperators accounting for cavity and atomic loss channels, respectively. The cavity Linblad term is ${\cal L}_\mathrm{cav} \hat \rho=(\kappa/2) {\cal D}[\hat a] \hat \rho$, $\kappa$ being the cavity loss rate, and 
\begin{equation}
{\cal D}[\hat a] \hat \rho=2\hat a \hat \rho \hat a^\dagger-\hat a^\dagger \hat a \hat \rho-\hat \rho \hat a^\dagger \hat a
\end{equation}
is the so-called Lindblad dissipator. Equation \eqref{eq:master} can be also recast in a the compact form $\dot{\hat \rho} = {\cal L} [\hat \rho]$, where ${\cal L}$ is the global Liouvillian superoperator~\cite{PetruccioneOQS}.
Although the atomic decay losses can also be modeled with jump operators, here, for simplicity, they will be added phenomenologically to the equations of motion of the reduced atomic density matrix later on. 

\begin{figure}[t!]
\centering
\includegraphics[trim=0.cm 12cm 0cm 2.cm,clip=true,scale=0.5,angle=0,origin=c]{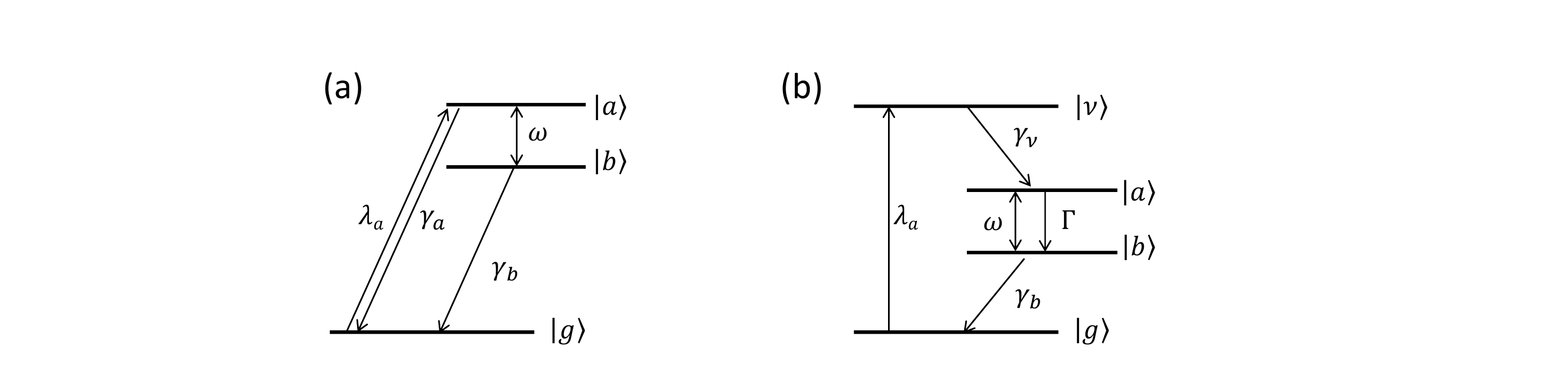}
\caption{(a) Effective three-level system for the class-A laser of Ref. \cite{scully1997quantum}. (b) Four-level system for the class-B laser.}
\label{fig:levels}
\end{figure}

First of all, we are interested in the reduced density matrix for photons $\hat \rho_{\mathrm{ph}}=\mathrm{Tr_{atoms}}[\hat \rho]$, whose matrix elements are noted $\rho_{nn'}\equiv\langle n| \hat \rho_{\mathrm{ph}} | n'\rangle$. The time evolution of the photonic reduced density matrix is, in the number Fock basis $\{ |n\rangle \}$, $n \in \mathbb{N} $  \cite{scully1997quantum}:
\begin{equation}
\dot { \rho} _{nn'}=-\frac{i}{\hbar} \mathrm{Tr}_{\mathrm{atoms}} [\hat{\mathscr{V}},\hat \rho]_{n,n'}+(  {\cal L}_{\mathrm{cav }} \hat \rho_{\mathrm{ph}})_{n,n'}.
\label{eq:eqmotion}
\end{equation}
The first term in Eq. \eqref{eq:eqmotion} is the atomic gain, which can be written as:
\begin{equation}
\left( \frac{\partial {\rho} _{nn'}}{\partial t} \right)_{\mathrm{gain}} =-\frac{i}{\hbar} \sum_{\{ \alpha \}} [\hat{\mathscr{V}},\hat \rho]_{\{ \alpha \},n;\{ \alpha \},n'},
\label{eq:gain}
\end{equation}
where $\{ \alpha \}=\alpha_1,\alpha_2,...,\alpha_N$, with the atomic-level index $\alpha_i=a$, $b$ or $g$, is the state of each of the N atoms. We consider the reduced density matrix of the $i$th atom plus the photonic field, as obtained by tracing out the remaining atomic degrees of freedom:
\begin{equation}
\hat \rho^i=\sum_{\{ \alpha_i' \}} \hat \rho_{\{ \alpha_i' \};\{ \alpha_i' \}},
\end{equation}
where $\{ \alpha_i' \}=\alpha_1,\alpha_2,...,\alpha_{i-1},\alpha_{i+1},...,\alpha_N$. Then Eq. \eqref{eq:gain} becomes
\begin{equation}
\begin{aligned}
\left( \frac{\partial {\rho} _{nn'}}{\partial t} \right)_{\mathrm{gain}}=& -\frac{i}{\hbar}  \sum_{i=1}^N \left( \mathscr{V}^{\vphantom{\prime}}_{a,n;b,n+1} \rho^i_{b,n+1;a,n'} - \rho^i_{a,n;b,n'+1} \mathscr{V}^{\vphantom{\prime}}_{b,n'+1;an'} \right.  \\  & +  \left. \mathscr{V}^{\vphantom{\prime}}_{b,n;a,n-1} \rho^i_{a,n-1;b,n'} - \rho^i_{b,n;a,n'-1} \mathscr{V}^{\vphantom{\prime}}_{a,n'-1;b,n'}  \right).
\end{aligned}
\label{eq:gainexplicit}
\end{equation}
So far we reproduced the results from Ref. \cite{scully1997quantum}. In order to make $N$ appear explicitly, we define the {\it normalized} generalized populations as 
\begin{equation}
 \rho_{\alpha,n;\beta, n'}=\frac{1}{N} \sum_{i=1}^N \rho^i_{\alpha, n;\beta,n'},
 \label{eq:generalized}
\end{equation}
which, unlike Ref. \cite{scully1997quantum}, has a prefactor $1/N$, ensuring it satisfies the density-matrix trace-one property. As a consequence, the number of atoms appears in the gain expression explicitly: 
\begin{equation}
\begin{aligned}
\left( \frac{\partial { \rho} _{nn'}}{\partial t} \right)_{\mathrm{gain}}=& -\frac{i N}{\hbar} \left( \mathscr{V}_{a,n;b,n+1} \rho_{b,n+1;a,n'} - \rho_{a,n;b,n'+1} \mathscr{V}_{b,n'+1;a,n'} \right.  \\  & +  \left. \mathscr{V}_{b,n;a,n-1} \rho_{a,n-1;b,n'} - \rho_{b,n;a,n'-1} \mathscr{V}_{a,n'-1;b,n'}  \right).
\end{aligned}
\label{eq:normgain}
\end{equation}
Equation \eqref{eq:normgain} shows that the total gain in a laser medium extensively scales with the number of atoms.

Secondly, in order to compute $\rho_{\alpha,n;\beta,n'}$, we are interested in the time evolution of the reduced atomic density matrix elements, governed by Eq. \eqref{eq:master}. The equation of motion can then be obtained by tracing out the master equation over all atoms $j \neq i$: 
\begin{equation}
\begin{aligned}
\dot{ \rho}^i_{\alpha,n;\beta,n'}   =-\frac{i}{\hbar} \langle  \alpha_i,n|  [\hat{\mathscr{V}^i},\hat \rho^i] | \beta_i,n' \rangle-\frac{i}{\hbar} \sum_{j \neq i} \langle  \alpha_i,n|  \sum_{\{ \alpha'_i \}}[\hat{\mathscr{V}^j},\hat \rho]_{\{ \alpha'_i \};\{ \alpha'_i \}} | \beta_i,n' \rangle \\
+ \left( \kappa \sqrt{(n+1)(n'+1)}  \rho^i_{\alpha,n+1;\beta,n'+1}- \kappa \sqrt{n n'}  \rho^i_{\alpha,n;\beta,n'} \right) +\text{atomic decay/pump terms}.
\end{aligned}
\label{eq:classArho}
\end{equation}

The first term in Eq. \eqref{eq:classArho} stands for single-atom contibutions. Keeping only such a term in the gain is equivalent to consider ---quoting Ref. \cite{scully1997quantum}--- \lq\lq that the atom-field population matrix for the $i$th atom may be treated as if the effect of other atoms is felt only through their contributions to $\rho_{nn'}$''. However there are two additional important contributions: the second and third terms of Eq. \eqref{eq:classArho}, which respectively account for two-atom quantum correlations ---the state of the $i$th-atom depends on the states of all other $j \neq i $ atoms---, and cavity dissipation. These two contributions have been neglected in Ref. \cite{scully1997quantum}; such an approximation leads to a closed system of equations, where the atomic variables  $\rho^i_{\alpha,n;\beta,n'}$ are adiabatically eliminated, hence solved in the steady state. The adiabatic elimination is justified in the context of class-A lasers because the cavity dissipation rate is assumed much smaller compared to the atomic decay rates, $\kappa\ll \gamma_{a,b}$. Consequently, analytical solutions for $\rho_{\alpha,n;\beta,n'}$ ---independent of the dissipation--- can be found and injected into Eq. \eqref{eq:eqmotion}. The latter can finally be analytically solved in the steady state. For completeness, we recall the result for the probability of emission of $n$ photons, $P_n \equiv \rho_{nn}$, in a class-A laser:
\begin{equation}
P_n= P_0 \prod _{k=1}^n \frac{\mathscr{A/\kappa}}{1+\frac{\mathscr{B}}{\mathscr{A}}\kappa}=P_0 \frac{\left( \frac{\mathscr{A}}{\mathscr{B}} \right) ! \left( \frac{\mathscr{A}^2}{\mathscr{B} \kappa} \right)^n}{\left( n+\frac{\mathscr{A}}{\mathscr{B}} \right) ! },
\label{eq:classAexact}
\end{equation}
where $\mathscr{A}=2r_a g^2/\gamma^2$ and  $\mathscr{B}=4 g^2 \mathscr{A}/\gamma^2$ are the linear gain and self-saturation coefficients, respectively, $P_0$ is determined by the normalization condition $\sum_n P_n=1$, and $r_a=\gamma \lambda_a/(\gamma+\lambda_a)$, upon assuming $\gamma_a=\gamma_b\equiv \gamma$. 

\subsection{Self-consistency}

The procedure reviewed above, leading to analytical results of the master equation for a class-A laser, represented an important milestone in laser physics theory. For the first time it was possible to describe not only the photon statistical properties of a laser in a quantum framework, but also its coherence properties, since the non-diagonal elements of the photonic density matrix are directly related to the spectral properties fo the emission. The latter, in particular, enabled the derivation of the Shallow-Townes relation from a quantum theory of light-matter interaction in an optical cavity. 

Here, we would like to discuss the self-consistency of this approach. Both equations governing the time evolution of photonic [Eq. \eqref{eq:eqmotion}] and atomic [Eq. \eqref{eq:classArho}] quantum states originate from the same master equation [Eq. \eqref{eq:master}]. Importantly, although neglecting the second and third terms in Eq. \eqref{eq:classArho} results in a good approximation for class-A laser systems, it is not justified in the general case, especially in class-B lasers. The self-consistency relations can be found from the identity
\begin{equation}
\rho_{nn'}=\rho^i_{a,n;a,n'}+\rho^i_{b,n;b,n'}+\rho^i_{g,n;g,n'}.
\label{eq:identity}
\end{equation}
This corresponds to the partial trace over the $i$th atom degrees of freedom.
Upon averaging both sides of Eq. \eqref{eq:identity} over the $N$ atoms, we obtain 
\begin{equation}
\rho_{nn'}=\rho_{a,n;a,n'}+\rho_{b,n;b,n'}+\rho_{g,n;g,n'},
\label{eq:identitynorm}
\end{equation}
where the terms in r.h.s. are now normalized generalized populations [Eq. \eqref{eq:generalized}]. Note that, in the absence of a factor $1/N$ normalizing the populations, Eq. \eqref{eq:identity} would have rather yielded $N\rho_{nn'}=\tilde \rho_{a,n;a,n'}+\tilde \rho_{b,n;b,n'}+\tilde \rho_{g,n;g,n'}$, where the tilde stands for unnormalized generalized populations. Therefore, for the sake of self-consistency, the linear gain coefficient $\mathscr{A}$ in Ref. \cite{scully1997quantum} should be multiplied by $N$. This agrees with our intuition that the gain scales with the number of atoms, which appears explicitly with our choice of normalized populations.

Moreover, differentiating Eq. \eqref{eq:identity} with respect to time and tracing over the states of the $i$th atom, we obtain 
\begin{equation}
\sum_{\alpha=a,b,g} \dot{\rho}_{\alpha,n;\alpha,n'}^i =\dot{\rho}_{nn'}.
\label{eq:consistency}
\end{equation}
Therefore, the r.h.s. of Eq. \eqref{eq:consistency} given by Eq. \eqref{eq:eqmotion} must be equal to the l.h.s. obtained from Eq. \eqref{eq:classArho}.  The self-consistency condition then reads:
\begin{equation}
\mathrm{Tr}_{\mathrm{i-at}} [ \dot{\hat \rho}^i_{nn'} ]=-\frac{i}{\hbar} \mathrm{Tr}_{\mathrm{atoms}} [\hat{\mathscr{V}},\hat \rho]_{nn'}+({\cal L}_\mathrm{cav}\hat \rho_{\mathrm{ph}})_{nn'}.
\label{eq:selfconsistency}
\end{equation}
where the r.h.s. is computed using the equation of motion of the the photonic reduced density matrix [Eq. \eqref{eq:eqmotion}], whereas the l.h.s. is computed using the one for the atomic reduced density matrix [Eq. \eqref{eq:classArho}].

Let us focus on the light-matter (gain) terms of the r.h.s. of  Eq \eqref{eq:selfconsistency}. From Eq. \eqref{eq:classArho}, they read: 
\begin{eqnarray}
\label{eq:proof1}
\left( \sum_{\alpha=a,b,g} \dot{\rho}_{\alpha,n;\alpha,n'}^i \right)_{\mathrm{gain}}  &=& -\frac{i}{\hbar} \sum_{\alpha_i=a,b,g} \langle  \alpha_i,n|  [\hat{\mathscr{V}^i},\hat \rho^i] | \alpha_i,n' \rangle-\frac{i}{\hbar} \sum_{j \neq i} \sum_{\alpha_j=a,b,g}\langle  \alpha_j,n|  [\hat{\mathscr{V}^j},\hat \rho^j] | \alpha_j,n' \rangle \\
&=& -\frac{i}{\hbar} \sum_{i=1}^N \mathrm{Tr}_{\mathrm{i-at}}  [\hat{\mathscr{V}^i},\hat \rho^i]_{nn'},
\label{eq:proof2}
\end{eqnarray}
thus we retrieve the r.h.s. of Eq. \eqref{eq:gainexplicit}---the gain. Note that this is possible thanks to the second term of Eq. \eqref{eq:proof1}: the two-atom contributions. These scale as $(N-1)$, therefore they cannot be neglected {\it a priori}, especially for a large number of atoms, which is the usual case in standard laser systems ---even at the microscale; they vanish only in cavities containing a single atom. Consequently, in general, both two-atom correlations and cavity dissipation in the equation of motion of the atomic reduced density matrix ---second and third terms in Eq. \eqref{eq:classArho}, respectively--- are necessary to ensure consistency of this approach. 

\subsection{Class-A laser threshold}
\label{Class-A threshold}

For simplicity we consider the same decay rate for atomic populations, $\gamma_a=\gamma_b\equiv \gamma$. Furthermore, we add pure dephasing effects with rate $\gamma_h$ ---not considered in Ref \cite{scully1997quantum}---, which govern the damping of the non-diagonal atomic density matrix elements. Note that $\gamma_h$ accounts for the homogeneous broadening of the gain and it is relevant for room temperature laser operation (weak coupling), where $\gamma_h\gg \gamma$. As concluded from the previous paragraphs, for the sake of consistency, we amend the linear gain coefficient as $\mathscr{A}\rightarrow N\mathscr{A}=2r_a g^2N/\gamma\gamma_h$, $\mathscr{B}\rightarrow4g^2\mathscr{A}/\gamma\gamma_h$. 
The laser threshold can be obtained from a detailed balance condition. As a matter of fact, the steady state of Eq. \eqref{eq:eqmotion} can be recast in four terms involving $P_{n-1}$, $P_n$, and $P_{n+1}$. Because of detailed balance, these terms cancel out two-by-two; the detailed balance condition reads: 
\begin{equation}
N \frac{4g^2 r_a}{4g^2n+\gamma \gamma_h} nP_{n-1}= 2\kappa n P_n.
\label{eq:classAdetailed}
\end{equation}
In Ref.  \cite{scully1997quantum}, the laser threshold has been derived within the linear gain approximation: neglecting the saturation term in the denominator of Eq. \ref{eq:classAdetailed}, the threshold pumping rate $\lambda^{sc}_{a,th}$ is given by 
\begin{equation}
N \frac{\lambda^{sc}_{a,th}}{1+\lambda^{sc}_{a,th}/\gamma}=\frac{\kappa\gamma_h}{2g^2}\gamma.
\label{eq:classAthreshold_sc}
\end{equation}
The r.h.s. of Eq. \ref{eq:classAthreshold_sc} is called the \lq\lq semiclassical threshold'' in Ref. \cite{PhysRevA.50.4318}, and it is valid as long as $4g^2\ll \gamma\gamma_h$. 
As we will see in Sec. \ref{Class-B threshold}, $\kappa\gamma_h/2g^2$ is the saturated population inversion above laser threshold.  
The factor $1/(1+\lambda^{sc}_{a,th}/\gamma)$ in the l.h.s. is a saturation function due to the fact that the maximum excitation rate per atom is $\gamma$. From the exact solution given in Eq. \eqref{eq:classAexact}, it is easy to show that such a 
threshold condition corresponds to a peak of the photon distribution at $n=0$. 

Here we rather introduce a generic definition of the laser threshold, valid independently of the details of the model considered and beyond the semiclassical approximation. Below the laser threshold the photon distribution is close to a thermal one, which is a monotonically decreasing function, $P_0>P_1$. In contrast, above the laser threshold, the photon distribution tends to a Poissonian distribution that has a single maximum, therefore $P_0<P_1$. The laser threshold is thus given by $P_0=P_1$ which, using Eq. \ref{eq:classAdetailed}, yields:
\begin{equation}
N \frac{\lambda_{a,th}}{1+\lambda_{a,th}/\gamma}=\frac{2\kappa}{ \beta_0},
\label{eq:classAthreshold}
\end{equation}
where $\beta_0$ stands for the  standard definition of the spontaneous emission factor (usually noted as $\beta$), 
\begin{equation}
\beta_0=\frac{4g^2}{4g^2+\gamma \gamma_h}, 
\label{eq:classA_beta} 
\end{equation}
whose inverse is the saturation photon number, $n_{sat}=\beta_0^{-1}$ \cite{PhysRevA.50.4318}. Importantly, the pumping rate at threshold given by the r.h.s. of Eq. \ref{eq:classAthreshold}, $2\kappa/\beta_0$, is in agreement with the definition of the laser threshold $P_ {thr}$ of Ref. \cite{PhysRevA.50.4318}. 
Therefore, our definition of the laser threshold as $P_0=P_1$ is in agreement with the rate equation analysis of Ref. \cite{PhysRevA.50.4318}, and only coincides with the semiclassical threshold in the thermodynamic limit ($n_{sat}\gg1$ or, equivalently, $4g^2\ll \gamma\gamma_h$).

Clearly, a single-atom class-A cavity may operate in the laser regime in the large photon number ---or small-$\beta$--- limit since $\kappa\ll \gamma $. 
However, as we will discuss in the next Section, a class-B laser ($\kappa > \gamma$) cannot operate above threshold in the large photon number limit unless the number of atoms is large enough, $N\gg 1$. In such a case then, both dissipation and $N$ are very large, therefore neither the two-atom contributions nor the cavity-dissipation in the reduced atomic density matrix can be neglected. 

\section{Class-B lasers}
\label{Class-B lasers}

In this section we derive a density matrix model for class-B lasers, for arbitrary $\beta$-factors and number of atoms $N$. Unlike the class-A model where the dynamics is governed by just one equation of motion for the photonic reduced density matrix, for class-B lasers we will come up with two nonlinear coupled equations, one for the photonic reduced density matrix, and the other one for the atomic reduced density matrix. This system of equations, even though it has no available analytical solution in general, is relatively simple and can be integrated with standard numerical methods such as the fourth-order Runge-Kutta algorithm. 

We recall that, in a class-B laser, the cavity dissipation rate is large compared to the atomic relaxation, $\kappa>\gamma$. In addition, class-B generally applies to room temperature operating devices, where pure dephasing $\gamma_h$ is very large. Hence, class-B lasers are defined as $\gamma_h \gg \kappa > \gamma$, and the only variables that can be adiabatically eliminated will be the non-diagonal elements of the atomic reduced density matrix. 

We will consider a 4-level atomic system, where the upper level is the pumping level $|\nu\rangle$, the upper and lower laser levels are $|a\rangle$ and $|b\rangle$, and the ground state is $|g\rangle$ (see Fig. 1). As a matter of fact, a four-level system is more general than a three-level one because most solid-state gain media are pumped above the upper laser state, such as conduction-band levels for electrons in the wetting layer of semiconductor quantum-dot lasers, or high-energy atomic levels in ion-doped crystals. The decay and pumping rates are defined as follows: $\gamma_\nu$ is the non-radiative decay rate from the pump to the upper laser level, $\Gamma$ is the spontaneous emission rate from $|a\rangle$ to $|b\rangle$, and $\gamma_b$ is the non-radiative decay rate from the lower laser level to the ground state. For further simplification, we will consider fast non-radiative decays, therefore will will eventually take the limit $\gamma_{\nu},\gamma_b\rightarrow+ \infty$. This approximation, which leads to unpopulated $|\nu\rangle$ and $|b\rangle$ states, is strictly valid for some class-B lasers such as certain rare earth-doped crystals (e.g. Nd:YAG), and leads to a minimal two-variable model. Yet, this simple model could be extended to other devices such as semiconductor (micro/nano) lasers by adding a thermal population to the lower laser state playing the role of the transparency carrier population, but the main model properties would remain unaffected.

\subsection{Atomic reduced density matrix}

As discussed in the previous section, two main equations of motion can be obtained from the master equation, Eq. \eqref{eq:master}. The time-domain differential equation for the photonic reduced density matrix is given by Eq. \eqref{eq:eqmotion}, while the one for the atomic reduced density matrix is given by Eq. \eqref{eq:classArho}. Unlike class-A, for class-B lasers it is not possible to neglect the two atom contributions and the dissipation in Eq. \eqref{eq:classArho}. However, an important simplification ---still legitimate in class-B lasers--- is the adiabatic elimination of the non-diagonal elements of $\rho^i_{\alpha n,\beta n'}$ ($\alpha\neq \beta$, $\alpha=a$ or $b$), since $\gamma_h \gg \gamma,\kappa$.  The dynamical equations read: 
\begin{equation}
\dot{ \rho}^i_{\alpha n,\beta n'}   =-\gamma_h \rho^i_{\alpha n,\beta n'} -\frac{i}{\hbar} \langle  \alpha_i,n|  [\hat{\mathscr{V}^i},\hat \rho^i] | \beta_i,n' \rangle+{\cal F}^i_{\alpha n,\beta n'}+{\cal G}^i_{\alpha n,\beta n'}
\label{eq:nondiagonal}
\end{equation}
where $(\alpha,\beta)=(a,b)$ or $(\alpha,\beta)=(b,a)$, ${\cal F}$ stands for \lq\lq two-atom contributions'' and ${\cal G}$ for dissipation (second and third terms of Eq. \eqref{eq:classArho}, respectively). Since $\gamma_h\gg \kappa$, ${\cal G}$ are subleading terms in Eq. \eqref{eq:nondiagonal}. Also, we neglect ${\cal F}$ terms in this equation, which will be justified in Sec. \ref{sec:conditional}. We now take the limit $\gamma_\nu, \gamma_b \rightarrow +\infty$, which yields $\rho^i_\nu, \rho^i_b\rightarrow 0$, and we adiabatically eliminate $\rho^i_{an,bn'+1}$ and $\rho^i_{bn+1,an'}$, which read: 
\begin{eqnarray}
\label{eq:adiabatic1}
\rho^i_{an,bn'+1}=\frac{i}{\hbar \gamma_h} \rho^i_{an,an'} \mathscr{V}_{an',bn'+1}=\frac{ig}{\hbar \gamma_h} \rho^i_{an,an'} \sqrt{n'+1} \\
\label{eq:adiabatic2}
\rho^i_{bn+1,an'}=-\frac{i}{\hbar \gamma_h} \rho^i_{an,an'} \mathscr{V}_{bn+1,an}=-\frac{ig}{\hbar \gamma_h} \rho^i_{an,an'} \sqrt{n+1},
\end{eqnarray}
where we have used $\mathscr{V}_{an',bn'+1}=\hbar g\sqrt{n'+1}$ and $\mathscr{V}_{bn+1,an}=\hbar g \sqrt{n+1}$, obtained from Eq. \eqref{eq:Hamiltonian}. 

We now compute the diagonal elements of Eq. \eqref{eq:classArho}, i.e. $\alpha_i=\beta_i=\alpha$, $\alpha=a,b$ or $g$. The ${\cal F}$-terms read: 
\begin{eqnarray}
{\cal F}^i_{\alpha n,\alpha n'}   &=& -\frac{i}{\hbar} \sum_{j \neq i} \langle  \alpha_i,n|  \sum_{\{ \alpha'_i \}}[\hat{\mathscr{V}^j},\hat \rho]_{\{ \alpha'_i \},\{ \alpha'_i \}} | \alpha_i,n' \rangle\label{eq:TAC1} \\
&=& \frac{i}{\hbar} \sum_{j \neq i} \langle  \alpha_i,n|  \sum_{ \alpha_j }\langle \alpha_j| [\hat{\mathscr{V}^j},\hat \rho^{ij}]|\alpha_j\rangle | \alpha_i,n' \rangle.
\label{eq:TAC2}
\end{eqnarray}
In Eq. \eqref{eq:TAC2} we have defined the two-atom reduced density matrix $\hat \rho^{ij}$ as follows
\begin{equation}
\hat \rho^{ij}=\sum_{\{  \alpha_{ij}' \} } \hat \rho_{\{  \alpha_{ij}' \} ,\{  \alpha_{ij}' \}},
\label{eq:twoatomrho}
\end{equation}
where $\{  \alpha_{ij}' \}=\alpha_1,...,\alpha_{i-1},\alpha_{i+1},..., \alpha_{j-1},\alpha_{j+1},...,\alpha_N$. Equation \eqref{eq:TAC2} can be computed explicitely: 
\begin{equation}
\begin{aligned}
{\cal F}^i_{\alpha n,\alpha n'}= - \frac{i}{\hbar} &\sum_{j \neq i} \left( \mathscr{V}_{an,bn+1} \langle  \alpha_i=\alpha,\alpha_j=b,n+1|  \hat \rho^{ij}|\alpha_i=\alpha,\alpha_j=a,n' \rangle  \right. \\
&-   \langle  \alpha_i=\alpha,\alpha_j=a,n|  \hat \rho^{ij} |\alpha_i=\alpha,\alpha_j=b,n'+1 \rangle \mathscr{V}_{bn'+1,an'}  \\
&+ \mathscr{V}_{bn,an-1} \langle  \alpha_i=\alpha,\alpha_j=a,n-1|  \hat \rho^{ij}|\alpha_i=\alpha,\alpha_j=b,n' \rangle \\
&-  \left.  \langle  \alpha_i=\alpha,\alpha_j=b,n|  \hat \rho^{ij}|\alpha_i=\alpha,\alpha_j=a,n'-1 \rangle  \mathscr{V}_{an'-1,bn'} \right).
\end{aligned}
\label{eq:TAC3}
\end{equation}

Clearly, all the matrix elements between brackets in Eq. \eqref{eq:TAC3} are unknown. Deriving an equation of motion for $\hat \rho^{ij}$ would lead to three-atom contributions, and so on and so forth. In Sec. \ref{sec:conditional} we introduce an ansatz to close the system of equations in a self-consistent way. Finally, the dissipation terms read: 
\begin{equation}
{\cal G}^i_{\alpha n,\beta n'}=  \kappa \sqrt{(n+1)(n'+1)}  \rho^i_{\alpha n+1,\beta n'+1}- \kappa \sqrt{n n'}  \rho^i_{\alpha n,\beta n'}.
\label{eq:G}
\end{equation}

\subsection{Conditional probability ansatz}
\label{sec:conditional}
The chosen ansatz must be such to decompose $\hat \rho^{ij}$ in terms of $\hat \rho^i$ and $\hat \rho^j$, under the constraint that it must verify the self-consistency condition [Eq. \eqref{eq:selfconsistency}]. 
The first intuitive choice would be full separability, $ \hat\rho =  \hat{\rho}_{\mathrm{ph}}\otimes \mathrm{Tr_{ph}}[\hat \rho^1] \otimes ... \otimes \mathrm{Tr_{ph}}[\hat \rho^N]$. 
This assumption, however, breaks the extensive scaling of the gain. 
Indeed, upon substituting the separable ansatz into Eq.~\eqref{eq:TAC1}, one obtains
\begin{align}
    \mathcal{F}^i_{\alpha n,\alpha n'} \overset{!}{=} -\frac{i}{\hbar} \langle  \alpha_i,n| \hat{\rho}^i | \alpha_i,n' \rangle \sum_{j \neq i} \mathrm{Tr}\lbrace[\hat{\mathscr{V}^j},\hat{\rho}^j]\rbrace = 0,
\end{align}
where the last equality trivially follows from the cyclic-permutation invariance of the trace.
An important conclusion is that, in a class-B laser, the density matrix is not factorizable, two-atom and photon-atom correlations are necessary for self consistency.

Our ansatz will be the product of a conditional probability for the $i$th atom and the reduced density matrix for the $j$th atom. We start by noticing that $\langle \alpha,n| \hat \rho^i |\alpha,n\rangle$ is the probability of finding the $i$th-atom in the state $|\alpha\rangle$ with $n$ photons. By definition, the conditional probability $P_i(\alpha |n) $ of finding the atom in the state $|\alpha\rangle$  provided there are $n$ photons is $P_i(\alpha|n)= \langle \alpha,n| \hat \rho^i |\alpha,n\rangle/P_n$. For a state of two atoms $i$ and $j$, we define the conditional probability ansatz as [for $(\delta,\epsilon)=(a,b)$ or $(b,a)$]
\begin{equation}
 \langle  \alpha_i=\alpha,\alpha_j=\delta ,n|  \hat \rho^{ij}|\alpha_i=\alpha,\alpha_j=\epsilon,n' \rangle\simeq P_i(\alpha|n,n') \rho^j_{\delta n,\epsilon n'}
 \label{eq:ansatz}
\end{equation}
where $P_i(\alpha|n,n')$ is the conditional probability of finding the $i$th-atom in the state $|\alpha \rangle$ provided there are $n$ {\it or} $n'$ photons, hence: 
\begin{equation}
P_i(\alpha|n,n')= \frac{P_i(\alpha|n) P_n+P_i(\alpha|n') P_{n'}}{P_n+P_{n'}} =\frac{\rho^i_{\alpha n,\alpha n}+\rho^i_{\alpha n',\alpha n'}}{\rho_{nn}+\rho_{n'n'}}.
\label{eq:conditional}
\end{equation}

Using the ansatz in Eq. \eqref{eq:ansatz}, $\mathrm{Tr_{i-at}}[\hat \rho^{ij}]=\hat \rho^j$. We can now compute ${\cal F}^i_{\alpha n,\alpha n'}$: 
\begin{eqnarray}
{\cal F}^i_{\alpha n,\alpha n'}&= -\dfrac{i}{\hbar} \sum_{j\neq i}  \left( \mathscr{V}_{an,bn+1} P_i(\alpha|n+1,n') \rho^j_{bn+1,an'} - P_i(\alpha|n,n'+1) \rho^j_{an,bn'+1} \mathscr{V}_{bn'+1,an'} \right.  \nonumber \\
&+ \left. \mathscr{V}_{bn,an-1} P_i(\alpha|n-1,n') \rho^j_{an-1,bn'} - P_i(\alpha|n,n'-1) \rho^j_{bn,an'-1} \mathscr{V}_{an'-1,bn'}  \right) \label{eq:F1}\\
&= -\dfrac{i}{\hbar} \sum_{j\neq i} \left( \sqrt{n+1}\, \frac{\rho^i_{\alpha n^{\vphantom{\prime}}+1,\alpha n+1}+\rho^i_{\alpha n',\alpha n'}}{\rho_{n^{\vphantom{\prime}}+1,n+1}+\rho_{n',n'}} \rho^j_{bn+1,an'} - \sqrt{n'+1} \, \frac{\rho^i_{\alpha n^{\vphantom{\prime}},\alpha n}+\rho^i_{\alpha n'+1,\alpha n'+1}}{\rho_{n,^{\vphantom{\prime}}n}+\rho_{n'+1,n'+1}} \rho^j_{an,bn'+1}  \right.  \nonumber \\
&+ \left. \sqrt{n} \, \frac{\rho^i_{\alpha n^{\vphantom{\prime}}-1,\alpha n-1}+\rho^i_{\alpha n',\alpha n'}}{\rho_{n^{\vphantom{\prime}}-1,n-1}+\rho_{n',n'}} \rho^j_{an-1,bn'}   - \sqrt{n'} \frac{\rho^i_{\alpha n^{\vphantom{\prime}},\alpha n}+\rho^i_{\alpha n'-1,\alpha n'-1}}{\rho_{n^{\vphantom{\prime}},n}+\rho_{n'-1,n'-1}}  \rho^j_{bn,an'-1} \right), \label{eq:F2}
\end{eqnarray}
and the equations of motion for the atomic reduced density matrix read:
\begin{eqnarray}
 \dot {\rho}^i_{an,an'} &=& \lambda_a \rho_{n,n'}-(\lambda_a+\Gamma) {\rho}^i_{an,an'}  -\frac{i}{\hbar}  \left( \mathscr{V}_{an,bn+1} \rho^i_{bn+1,an'} - \rho^i_{an,bn'+1} \mathscr{V}_{bn'+1,an'} \right)+{\cal F}^i_{a n,a n'}+{\cal G}^i_{a n,a n'}  \label{eq:classBrhoa} \\
 \dot {\rho}^i_{bn,bn'} &=& \Gamma \rho_{an,an'}-\gamma_b {\rho}^i_{bn,bn'}  -\frac{i}{\hbar}  \left( \mathscr{V}_{bn,an-1} \rho^i_{an-1,bn'} - \rho^i_{bn,an'-1} \mathscr{V}_{an'-1,bn'} \right)+{\cal F}^i_{b n,b n'}+{\cal G}^i_{b n,b n'}  \label{eq:classBrhob} \\
 \dot {\rho}^i_{gn,gn'} &=& \gamma_b \rho_{bn,bn'}-\lambda_a \rho^i_{gn,gn'} +{\cal F}^i_{g n,g n'}+{\cal G}^i_{g n,g n'}  \label{eq:classBrhog}
\label{eq:}
\end{eqnarray}
where the ${\cal  G}$-terms are given by Eq. \eqref{eq:G}.

The two-atom contributions as resulting from the ansatz \eqref{eq:ansatz} verify the self-consistency relation \eqref{eq:selfconsistency}. Indeed, summing out Eq. \eqref{eq:F1} over the $i$th-atom states gives :
\begin{equation}
\sum_{\alpha=a,b,g} {\cal F}^i_{\alpha n,\alpha n'}=\sum_{j \neq i} \left( \mathscr{V}_{an,bn+1}  \rho^j_{bn+1,an'} -  \rho^j_{an,bn'+1} \mathscr{V}_{bn'+1,an'} + \mathscr{V}_{bn,an-1} \rho^j_{an-1,bn'} -  \rho^j_{bn,an'-1} \mathscr{V}_{an'-1,bn'}  \right),
\end{equation}
 and then 
\begin{eqnarray}
\label{eq:xxxxxxx}
 \sum_{\alpha=a,b,g} \dot{\rho}_{\alpha n,\alpha n'}^i &=& -\frac{i}{\hbar} \sum_{i=1}^N \mathrm{Tr}_{\mathrm{i-at}}  [\hat{\mathscr{V}^i},\hat \rho^i]_{n,n'} +({\cal L}_\mathrm{cav}\hat \rho_{\mathrm{ph}})_{n,n'},
\label{eq:proof2}
\end{eqnarray}
which is the equation of motion of the photonic reduced density matrix.  Consequently, the ansatz Eq. \eqref{eq:ansatz} allows us to close the system in a self-consistent way. Note that Eqs. \eqref{eq:classBrhoa}-\eqref{eq:classBrhog} are quite general in a three-level system ---the pumping level having been adiabatically eliminated---, regardless of whether $|b\rangle$ is also adiabatically eliminated or not. As stated before, our choice of $\gamma_b\rightarrow +\infty$ leads to $\rho_{bn,bn'} \rightarrow 0$, and therefore only Eq. \eqref{eq:classBrhoa} needs to be taken into account, considerably simplifying our system of equations since it will be eventually reduced to only two dynamical variables, $\rho_{an,an'}$ and $\rho_{nn'}$.

Finally, on the basis of this ansatz, we can justify to have neglected two-atom contributions in the adiabatic elimination leading to Eqs. \eqref{eq:adiabatic1}-\eqref{eq:adiabatic2}. Since $\alpha \neq \beta$, the bi-atomic corrections  ${\cal F}^i_{\alpha n,\beta n'}$ contain subleading terms such as $\rho^i_{\alpha n,\beta n}$. 

\subsection{Dynamical equations for the class-B density matrix approach}

The last step is to sum up Eq. \eqref{eq:classBrhoa} over $j$ and $i$. First of all, we perform the sum over $j\neq i$ in Eq. \eqref{eq:F1}:
\begin{eqnarray}
 -\dfrac{i}{\hbar} \left[ \mathscr{V}_{an,bn+1} P_i(\alpha|n+1,n') (N\rho_{bn+1,an'}-\rho^i_{bn+1,an'}) - P_i(\alpha|n,n'+1) (N\rho_{an,bn'+1}-\rho^i_{an,bn'+1}) \mathscr{V}_{bn'+1,an'} \right.  \nonumber \\
+ \left. \mathscr{V}_{bn,an-1} P_i(\alpha|n-1,n') (N\rho_{an-1,bn'}-\rho^i_{an-1,bn'}) - P_i(\alpha|n,n'-1) (N\rho_{bn,an'-1}-\rho_{bn,an'-1}) \mathscr{V}_{an'-1,bn'}  \right]. \label{eq:expression}
\end{eqnarray}
The second terms inside the four parenthesis are of order $O(1)$, and they can thus be neglected if $N\gg 1$. However, we will take them into account since we also want to describe cavities with few atoms. 
We then make the approximation of a spatially homogeneous medium, 
\begin{equation}
\sum_i P_i(\alpha|n,n')\rho^i_{\alpha n,\beta n'}\approx P(\alpha|n,n') \sum_i \rho^i_{\alpha n,\beta n'},
\end{equation}
where we have defined the spatially averaged conditional probability as $P(\alpha|n,n')=(1/N) \sum_i P_i(\alpha|n,n')=(\rho_{\alpha n,\alpha n}+\rho_{\alpha n',\alpha n'})/(\rho_{nn}+\rho_{n'n'})$. Summing up expression \eqref{eq:expression} over $i$ we obtain: 
\begin{eqnarray}
\frac{1}{N}\sum_i {\cal F}^i_{\alpha n,\alpha n'} &=
 -\dfrac{i}{\hbar} (N-1) \left[ \mathscr{V}_{an,bn+1} P_i(\alpha|n+1,n') \rho_{bn+1,an'} - P_i(\alpha|n,n'+1) \rho_{an,bn'+1} \mathscr{V}_{bn'+1,an'} \right.  \nonumber \\
&+ \left. \mathscr{V}_{bn,an-1} P_i(\alpha|n-1,n') \rho^i_{an-1,bn'} - P_i(\alpha|n,n'-1) \rho_{bn,an'-1} \mathscr{V}_{an'-1,bn'}  \right]. 
\label{eq:sumexpression}
\end{eqnarray}
Finally, summing Eq. \eqref{eq:classBrhoa} over $i$, and using Eqs. \eqref{eq:G} and \eqref{eq:sumexpression}, we obtain 
\begin{equation}
\begin{aligned}
 \dot {\rho}_{an,an'} &= \lambda_a \rho_{n,n'}-(\lambda_a+\Gamma) {\rho}_{an,an'}  -\frac{i}{\hbar}  \left( \mathscr{V}_{an,bn+1} \rho_{bn+1,an'} - \rho_{an,bn'+1} \mathscr{V}_{bn'+1,an'} \right) \nonumber \\
 &-\dfrac{i}{\hbar} (N-1) \left[ \mathscr{V}_{an,bn+1} \frac{\rho_{\alpha n^{\vphantom{\prime}}+1,\alpha n+1}+\rho_{\alpha n',\alpha n'}}{\rho_{n^{\vphantom{\prime}}+1,n+1}+\rho_{n',n'}} \rho_{bn+1,an'} -\frac{\rho_{\alpha n^{\vphantom{\prime}},\alpha n}+\rho_{\alpha n'+1,\alpha n'+1}}{\rho_{n,^{\vphantom{\prime}}n}+\rho_{n'+1,n'+1}} \rho_{an,bn'+1} \mathscr{V}_{bn'+1,an'} \right.  \nonumber \\
&+ \left. \mathscr{V}_{bn,an-1} \frac{\rho_{\alpha n^{\vphantom{\prime}}-1,\alpha n-1}+\rho_{\alpha n',\alpha n'}}{\rho_{n^{\vphantom{\prime}}-1,n-1}+\rho_{n',n'}} \rho_{an-1,bn'} - \frac{\rho_{\alpha n^{\vphantom{\prime}},\alpha n}+\rho_{\alpha n'-1,\alpha n'-1}}{\rho_{n^{\vphantom{\prime}},n}+\rho_{n'-1,n'-1}} \rho_{bn,an'-1} \mathscr{V}_{an'-1,bn'}  \right] \\
&+  \kappa \sqrt{(n+1)(n'+1)}  \rho_{a n+1,a n'+1}- \kappa \sqrt{n n'}  \rho_{a n,a n'}  .\label{eq:sumexpression1}
\end{aligned}
\end{equation}
The terms inside the brackets are the two-atom contributions. Clearly, they vanish when $N=1$, and the class-A-like model as in Ref. \cite{scully1997quantum} is retrieved. Yet, they cannot be neglected in general ---even in few atom cavities--- because they balance the dissipation terms, which are large in class-B laser systems. 

Using the shortcut notation for the diagonal reduced density matrix elements $\rho^a_n\equiv \rho^{\vphantom{\prime}}_{an,an}$ and $P^{\vphantom{\prime}}_n\equiv \rho^{\vphantom{\prime}}_{nn}$, the final form of the equations of motion of a class-B laser reads: 
\begin{eqnarray}
 \dot {\rho}^a_n &= \lambda_a P_n-\left[ \lambda_a+\Gamma+\frac{2g^2}{\gamma_h}(n+1) \right]  \rho^a_n +  \kappa (n+1) \rho^a_{n+1} - \kappa n  \rho^a_n  \nonumber \\
 &+\dfrac{2 g^2(N-1)}{\gamma_h}  \left[ n  \rho^a_{n-1} \, \dfrac{\rho^a_{n-1}+\rho^a_n}{P_{n-1}+P_n} -(n+1) \rho^a_n \, \dfrac{\rho^a_n+\rho^a_{n+1}}{P_n+P_{n+1}}  \right] \label{eq:modela}\\
 \dot {P}_n &= \dfrac{2g^2N}{\gamma_h}\left[ n \rho^a_{n-1}-(n+1)\rho^a_n \right]  +  \kappa (n+1) P_{n+1} - \kappa n  P_n.  \label{eq:modelP}
\end{eqnarray}
Equations \eqref{eq:modela}-\eqref{eq:modelP} represent a two-variable nonlinear dynamical system extended on a discrete pseudo-space dimension given by the photon number. 
They can be applied to any class-B laser system, from small to very large $N$, and for any spontaneous emission factor $\beta$, 
\begin{equation} 
\beta= \frac{2g^2}{2g^2+(\Gamma+\lambda_a)\gamma_h}.
\end{equation}
Note that, unlike Eq. \eqref{eq:classA_beta}, the $\beta$-factor in this four-level system depends on the pump rate, which comes from the fact that the spontaneous recombinations are due to transitions $|a\rangle \rightarrow |b\rangle$, instead of $|a\rangle, |b\rangle  \rightarrow |g\rangle$ used in the class-A model (Sec. \ref{Class-A lasers}). The former better describe spontaneous emission, whereas $|a\rangle, |b\rangle  \rightarrow |g\rangle$ rather model non-radiative recombinations. Therefore, we predict that the spontaneous emission factor to be pump-dependent in a quantum description, and such a dependency is enhanced for hard pumping, $\lambda_a>\Gamma$.

Our approach enables tackling classical and/or quantum fundamental properties of different laser systems, from very small ones or \lq\lq nanoscopic'' (few atoms, few photons), all the way to the thermodynamic limit (massively large photon and atomic numbers). 
It is usual to take the characteristic photon number in the resonators as a good thermodynamic parameter quantifying the \lq\lq system size''. This is given by the saturation photon number $n_{sat}=\beta^{-1}$. Three regimes can thus be distinguished for increasing typical photon numbers \cite{2015NatSR...515858W,PhysRevLett.124.213602}: $\beta^{-1} \lesssim 10^2$, $10^2 \lesssim \beta^{-1} \lesssim 10^4$, and $\beta^{-1} \gtrsim 10^4$, corresponding to the nanoscopic, mesoscopic, and macroscopic regimes, respectively. Needless to say, most of what we know of class-B lasers comes from a semiclassical description of macroscopic devices, which must be described by the macroscopic regime in the thermodynamic limit. On the opposite side, micro and nanolasers in the so-called \lq\lq thresholdless'' regime ($\beta\sim 1$) below to the nanoscopic world. Such a situation can be described with Eqs. \eqref{eq:modela}-\eqref{eq:modelP}. Importantly, also mesoscopic situations (hundreds of photons), which usually elude standard theoretical limits, can also be captured by this model; this is important from a technological point of view since many semiconductor nanolasers exhibit $\beta$-factors in the range $\sim 0.01-0.1$ therefore operating at the frontier between nanoscopic and mesoscopic regimes.   

\begin{figure}[h!]
\centering
\includegraphics[trim=0.cm 2cm 0cm 2.cm,clip=true,scale=0.5,angle=0,origin=c]{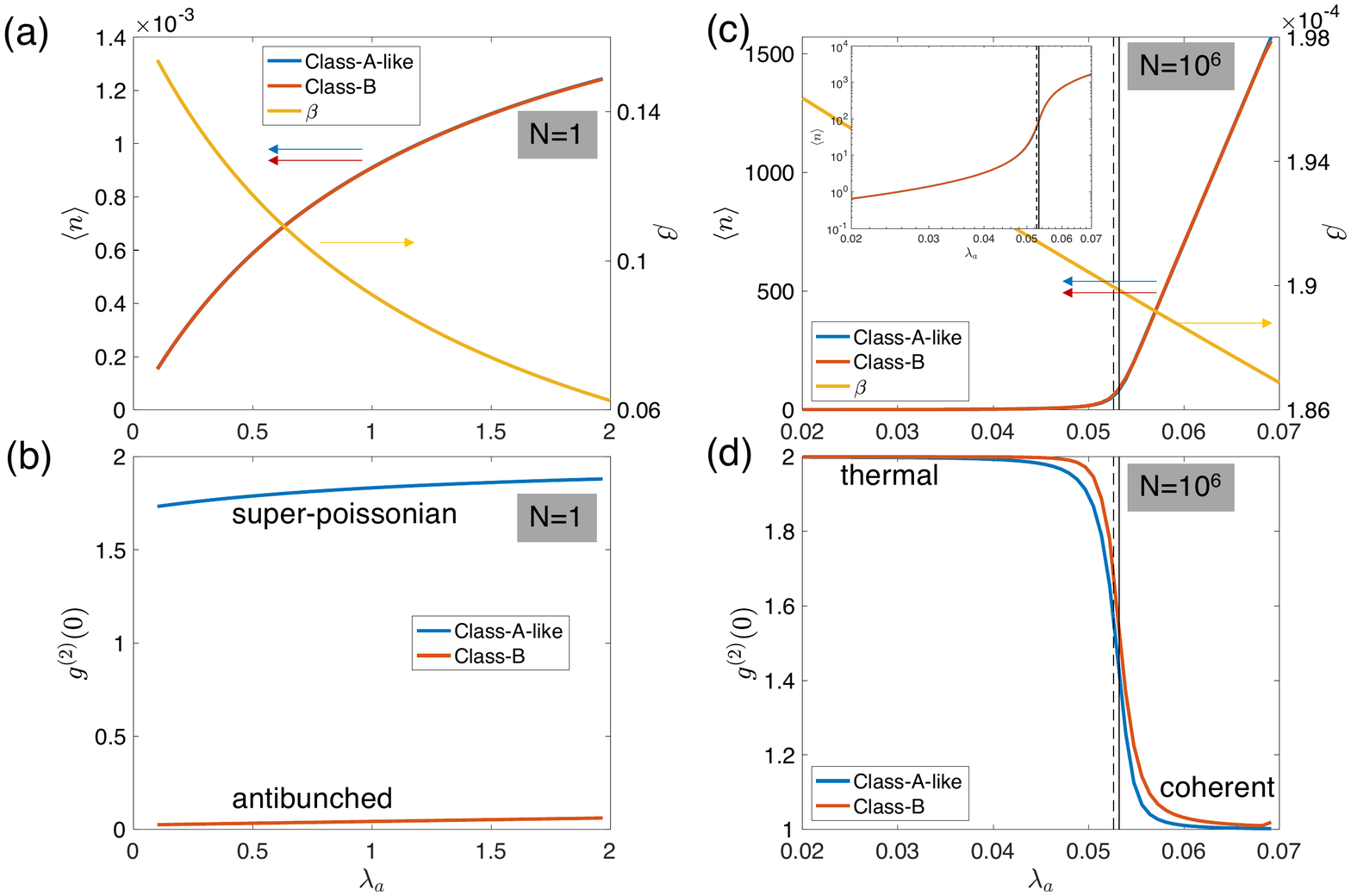}
\caption{Numerical integration of the quantum density matrix model for class-B regime (fourth-order Runge-Kutta simulations) . (a)-(b) Single atom cavity limit. Parameters are $\kappa=100$, $\Gamma=1$, $\gamma_h=10^3$, $g=10$ and $N=1$, cutoff of the Fock basis $N_{cutoff}=50$, and integration time step $h=10^{-5}$. (c)-(d) Towards the thermodynamic limit. Parameters are $\kappa=100$, $\gamma_h=10^4$, $g=1$ and $N=10^6$. The cutoff for the Fock-basis is $N_{cutoff}=50$ below threshold ($\lambda_a\leq \lambda^{(0)}_{a,th}$), and $N_{cutoff}=2200$ near and above threshold  ($\lambda_a>\lambda^{(0)}_{a,th}$); the integration time step is $h=2\times 10^{-6}$. $\lambda_a$ is the pumping rate. Total integration time $T=2 $. Time unit= $\Gamma^{-1}$ (equivalently, $\Gamma=1$ in the simulations).}
\label{fig:single-thermo}
\end{figure}

\subsection{Numerical simulations}
\label{Class-B numerics}

Let us start by considering two important strongly dissipative ($\kappa \gg \Gamma$) limits of Eqs. \eqref{eq:modela}-\eqref{eq:modelP}: the single atom cavities ($N=1$), and the thermodynamic limit, for which $N\gg 1$ and the saturation photon number is macroscopic ($n_{sat}=\beta^{-1}\gtrsim 10^4$). 
The latter admits a laser threshold, and encompasses the great majority of well-known solid-state lasers, from semiconductor edge-emitters to macroscopic ion-doped-crystal lasers such as Ti:Sa or Nd:YAG Fabry-Perot laser cavities and also Er:doped fiber lasers, among others. 

We first adress the single atom case, which already provides a striking difference with respect to the the class-A model predictions. A single-atom class-B device can only overcome laser threshold for large enough $N$, and \lq\lq few atom'' cavities can reach laser threshold only with large $\beta$-factors. Specifically, as we will explicitly derive in Sec. \ref{Class-B threshold}, a device can theoretically reach laser threshold provided $N$ is large enough, $N>\kappa \gamma_h/2g^2$. We numerically integrate Eqs. \eqref{eq:modela}-\eqref{eq:modelP} for $N=1$ and $\beta\sim 0.1$; the parameters of Fig. \ref{fig:single-thermo}a-b yield $\kappa \gamma_h/2g^2=500$ and therefore no laser-threshold can be attained with $N=1$. It is observed that the mean photon number in the class-B model is approximately the same as for the class-A-like counterpart (Fig. \ref{fig:single-thermo}a). Nevertheless, there is a strong qualitative difference in second order correlations (Fig. \ref{fig:single-thermo}b). While the class-A model only predicts super-poissonian light  [$g^{(2)}(0)>1$], the class-B model features strong antibunching [$g^{(2)}(0)\sim 0$]. This case is akin to a Jaynes Cummings density matrix model (see, e.g., Ref. \cite{PhysRevB.70.115304} for quantum dot excitons) with vanishing atom-cavity detuning and strong pure dephasing. 

On the other hand, in order to approach the thermodynamic limit we numerically integrate Eqs. \eqref{eq:modela}-\eqref{eq:modelP} for $N=10^{6}$ and $\beta\sim 10^{-4}$ (Fig. \ref{fig:single-thermo}c-d). The salient features of a second order phase transition are apparent, both as a sharp ---but continuous--- increase of the photon number at threshold, and a sharp decrease of $g^{(2)}(0)$ from $g^{(2)}(0)\approx 2$ (spontaneous emission) to $g^{(2)}(0)\approx 1$ (coherent laser emission). In this limit the diagonal elements of the class-B density matrix compare very well with those of the class-A-like model, which is translated as a very good agreement both in terms of mean photon number (Fig. \ref{fig:single-thermo}c) and second order correlations (Fig. \ref{fig:single-thermo}d).

\begin{figure}[h!]
\centering
\includegraphics[trim=0.cm 2cm 0cm 2.5cm,clip=true,scale=0.5,angle=0,origin=c]{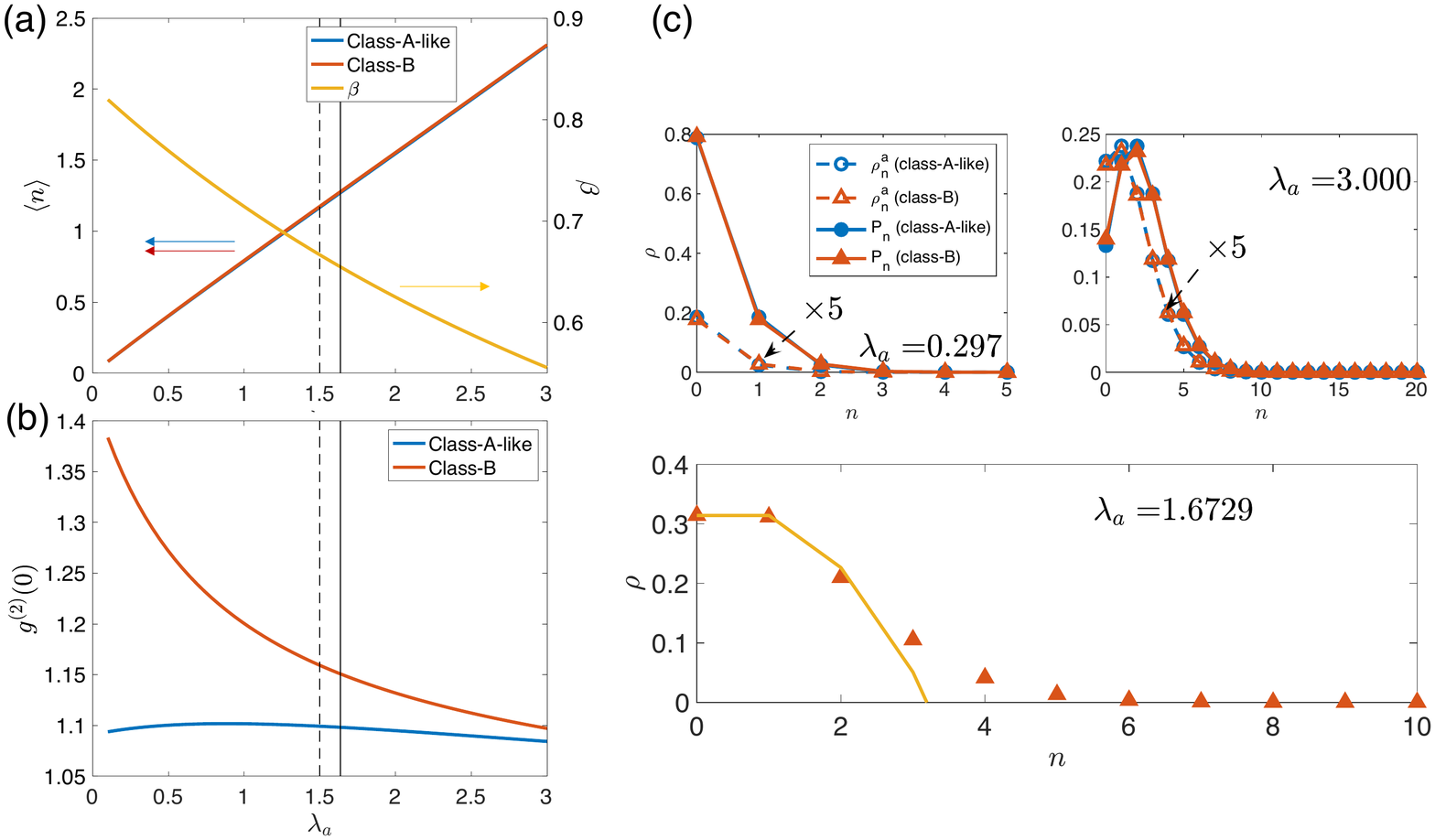}
\caption{Numerical integration of the quantum density matrix model for class-B laser emission: nanoscopic regime (few atoms and few photons). Parameters are $\kappa=10$, $\Gamma=1$, $\gamma_h=10^3$, $g=50$ and $N=10$. Fourth-order Runge-Kutta simualtions have been used, with time-step $h=10^{-5}$, and cutoff for the Fock-basis of  $N_{cutoff}=50$ below threshold ($\lambda_a\leq 3\lambda^{(0)}_{a,th}/4$), and $N_{cutoff}=100$ near and above threshold  ($\lambda_a>3\lambda^{(0)}_{a,th}/4$). (a) Mean photon number (left axis) and $\beta$-factor (yellow, right axis) as a function of the pump rate $\lambda_a$ for the class-A-like solutions (blue) and the class-B solutions (red). (b) Second order coherence. Dashed line marks the class-A-like threshold denoted $\lambda^{(0)}_{a,th}$ in the text, and solide line corresponds to the class-B threshold approximation, $\lambda^{(1)}_{th}$. (c) Photon number distributions (solid lines), and the upper-laser level atomic occupation probability (dashed lines) for three cases: below (top, left), above (top, right) and on-threshold (below, only $P_n$ class-B solution ---red line --- is displayed), defined as $\lambda_a=\lambda_{a,th}$ such that the (class-B) solution verifies $P_0\approx P_1$. Yellow line: approximate second order polynomial.}
\label{fig:nanoscopic}
\end{figure}

\begin{figure}[h!]
\centering
\includegraphics[trim=0.cm 2cm 0cm 2.cm,clip=true,scale=0.5,angle=0,origin=c]{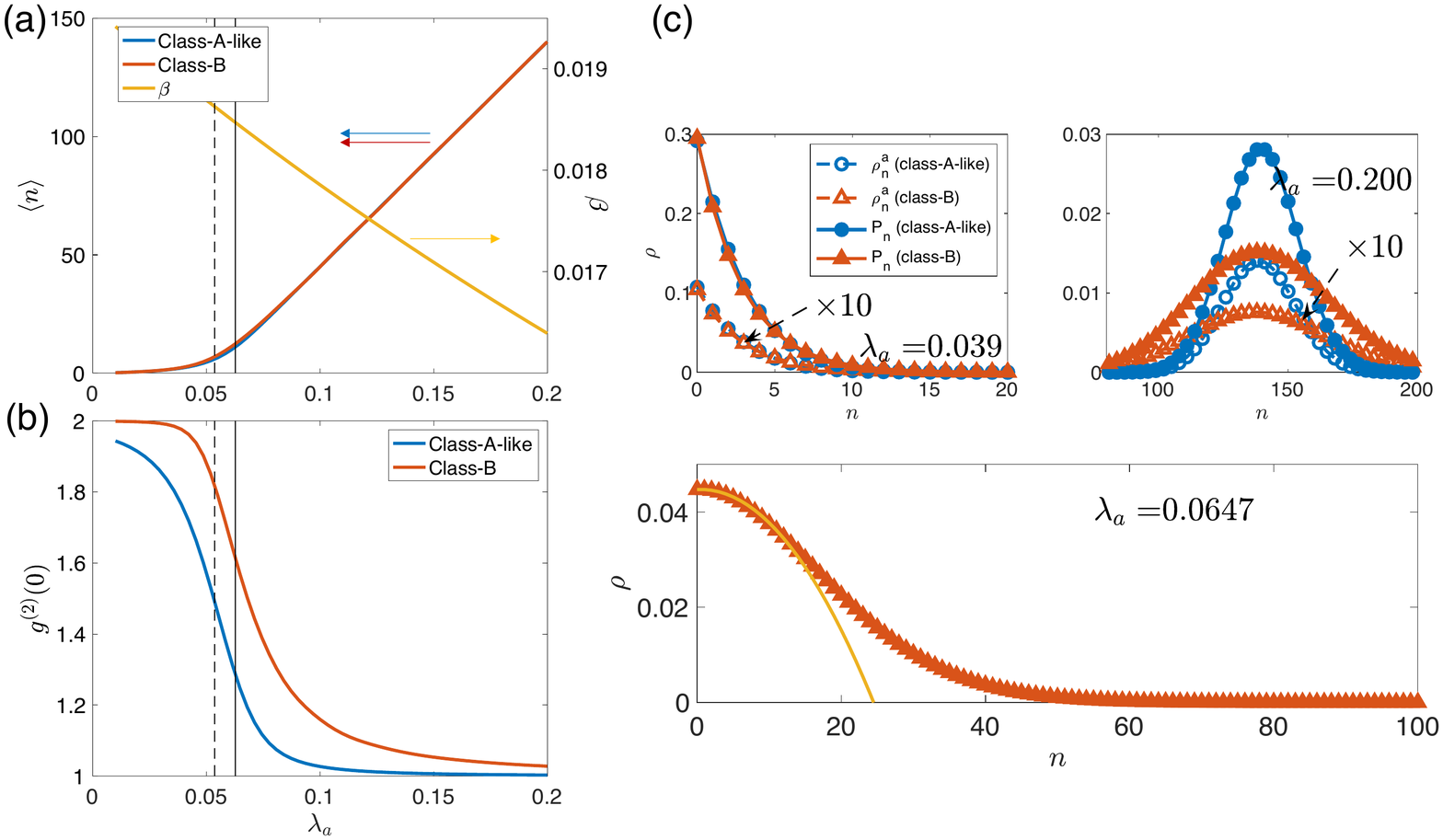}
\caption{Numerical integration of the quantum density matrix model for class-B laser emission: mesoscopic regime. Parameters are $\kappa=100$, $\Gamma=1$, $\gamma_h=10^4$, $g=10$ and $N=10^5$. Fourth-order Runge-Kutta simualtions have been used, with time-step $h=10^{-5}$, and cutoff for the Fock-basis of  $N_{cutoff}=50$ below threshold ($\lambda_a\leq 3\lambda^{(0)}_{a,th}/4$), and $N_{cutoff}=300$ near and above threshold  ($\lambda_a>3\lambda^{(0)}_{a,th}/4$). (a)-(c) Same as in Fig. \ref{fig:nanoscopic}.}
\label{fig:mesoscopic}
\end{figure}

In between these two limit cases, in general, deviations from class-A-like solutions are quite large in terms of second order correlations, while mean photon numbers are very close to each other when comparing class-A-like and class-B models. We illustrate this by numerically integrating two class-B lasing regimes: the nanoscopic ($\beta^{-1} \sim  1-2$, Fig. \ref{fig:nanoscopic}) and the mesoscopic ($\beta^{-1} \sim 100$, Fig. \ref{fig:mesoscopic}) regimes. In the mesoscopic case of Fig. \ref{fig:mesoscopic} there is a clear threshold ---reminiscent of the phase transition in the thermodynamic limit--, both  in the mean photon number and in $g^{(2)}(0)$. The parameters chosen for this mesoscopic case approximately represent a semcionductor nanolaser with embedded quantum wells, with $\beta$-factors of the order of $\beta\sim 10^{-2}$, Q-factors of a few thousand and optical volumes as small as $V\sim 0.1 \mu \mathrm{m}^3$, compatible with $N=10^5$ emitters in the cavity. In the nanoscopic regime of Fig. \ref{fig:nanoscopic}, on the other hand, there is no clear laser threshold in the dependence of the mean photon number with the pump rate---there is no signature of any phase transition. Only the second order correlations feature a transition towards coherence in the class-B-model, while the class-A-like model correlations remain approximately constant around $g^{(2)}(0)\sim 1.1$. 

\subsection{Class-B laser threshold}
\label{Class-B threshold}

A simple detailed balance condition can be deduced for the steady state of Eq. \eqref{eq:modelP}:
\begin{equation}
N \frac{2g^2}{\gamma_h} \rho^a_{n-1}= \kappa P_n,
\label{eq:classBdetailed}
\end{equation}
valid for $n\geq 1$. Summing up Eq. \eqref{eq:classBdetailed} leads to the steady state occupation probability of the upper laser level, $\mathrm{Tr_{ph}} [\hat \rho^a]=\kappa \gamma_h(1-P_0)/2Ng^2 $, and hence to the steady state population inversion,
\begin{equation}
 N_{ss}=\frac{\kappa \gamma_h (1-P_0)}{2g^2} .
\label{eq:Nth}
 \end{equation}
An important physical quantity is the saturated population inversion in the large photon number  limit, $N_{sat}=\kappa \gamma_h/2g^2$, which is the steady state population inversion for $\langle n \rangle \gg n_{sat}$, and hence $P_0\approx 0$. This equals the gain clamping to optical losses above the laser threshold within a mean-field theory, i. e., neglecting spontaneous emission \cite{PhysRevA.50.4318}.  
Notice however that, within the laser oscillation regime, the population inversion will clamp in general below $N_{sat}$, which can be interpreted as a lowering of steady state gain due to the spontaneous emission contribution to the lasing mode. 
  
Let us now derive the class-B laser threshold. In an otherwise class-A regime, because atom-atom correlations and dissipation can be neglected in  Eq. \eqref{eq:modela}, its steady state would simply lead to a relation between $\rho^a_n$ and $P_n$, and hence to a relation between $P_{n-1}$ and $P_n$ similar to Eq. \eqref{eq:classAdetailed}. 
Let us call $\lambda^{(0)}_{a,th}$ the resulting class-A-like laser threshold in the four-level system. The lasing condition introduced in Sec. \ref{Class-A threshold}, $P_0=P_1$, leads to
\begin{equation}
\lambda^{(0)}_{a,th}= \frac{\kappa}{N\beta}=\frac{\frac{N_{sat}}{N}\left( \Gamma+\frac{2g^2}{\gamma_h}\right)}{1-\frac{N_{sat}}{N}}.
\label{eq:lambda_th0}
\end{equation}

Unlike class-A lasers, for the full class-B system [Eqs. \eqref{eq:modela}-\eqref{eq:modelP}] there is no closed analytical form for the pumping rate at laser threshold. We then proceed perturbatively: 
\begin{equation}
\lambda_{a,th}=\lambda^{(0)}_{a,th}+\delta ^{(1)}+...\equiv \lambda^{(1)}_{a,th}+...
\label{classBth}
\end{equation}
where $\delta ^{(1)}$ is the leading order correction; $\delta ^{(1)}$ can be found by
approximating the photon probability distribution maximum by a quadratic polynomial at laser threshold for small $n$: 
\begin{equation}
P_n\Big|_{\lambda= \lambda_{a,th}}\approx P_0\left\{ 1+\ \xi \left[ \frac{1}{4}-\left(n-\frac{1}{2}\right) ^2 \right]\right\},
\label{eq:quadraticPn}
\end{equation}
where $\xi$ is a parameter to be determined self-consistently; note that Eq. \eqref{eq:quadraticPn} verifies $P_0=P_1$. Inserting Eqs. \eqref{eq:quadraticPn} and \eqref{eq:classBdetailed} into the steady state solution of Eq. \eqref{eq:modela} for $n=0$, we obtain:
\begin{equation}
\delta^{(1)}=\frac{\frac{N_{sat}}{N}\left[ \xi-\frac{(1-\xi)}{N}\right]}{1-\frac{N_{sat}}{N}}.
\label{Dlam1}
\end{equation}

An immediate consequence of Eq. \ref{classBth}, together with Eqs. \ref{eq:lambda_th0} and \ref{Dlam1}, is that a necessary condition for laser oscillation is $N>N_{sat}$. This is not obvious a priori because, as stated above, population inversion clamping generally occurs below $N_{sat}$. Furthermore, $\xi$ can be obtained by evaluating the the roots of the r.h.s. of Eq. \eqref{eq:modela} for $n=1$. In the limit $\xi \ll 1$ and $B\equiv N_{sat}/N\ll1$, $\xi$ becomes one of the following roots of a second order polynomial:
\begin{equation}
\xi_\pm=\frac{1}{2} \cdot \frac{5g^2 +4\kappa \gamma_h (N-1)/N+\Gamma\gamma_h -4 \kappa \gamma_h \pm \sqrt{\Delta}}{4g^2 +7\kappa \gamma_h(N-1)/N+\Gamma \gamma_h-4 \kappa \gamma_h},
\label{eq:xiroots}
\end{equation}
where $\Delta=-8g^4[2+7 B(N-1)]-4 g^2 (\Gamma-4 \kappa)\gamma_h+( g^2[5+8B(N-1)]+(\Gamma-4\kappa)\gamma_h ) ^2$.

First of all, two important limit cases are worth to be discussed, leading to real roots: $N=1$ and $N\gg N_{sat}$. For $N=1$, and in the weak dissipation limit ($\kappa \rightarrow 0$, the class-A Scully's limit), Eq. \eqref{eq:xiroots} leads to $\xi_-= 2g^2/(4g^2+\Gamma \gamma_h)$ and the threshold correction vanishes, 
$\delta ^{(1)} \rightarrow 0$, as expected. In the opposite ---thermodynamic--- limit ($N\gg N_{sat}$ and $\beta\rightarrow 0$) the threshold correction is also small, $\delta^{(1)}\approx  [\xi N_{sat}/N]/[1-N_{sat}/N]$ (see vertical lines in Fig. \ref{fig:single-thermo}c-d).

Interestingly, threshold corrections are larger from the nanoscopic (Fig. \ref{fig:nanoscopic}) to the mesoscopic (Fig. \ref{fig:mesoscopic}) class-B regimes where $\xi_\pm$ are complex. Complex $\xi_\pm$  can be attributed to errors due to the quadratic approximation of Eq. \eqref{eq:quadraticPn}; we then approximate $\xi \approx \operatorname{Re} (\xi_-)$ in those cases. In the nanoscopic regime, $\delta^{(1)}$ is less that 10\% with respect to the class-A-like threshold (Fig. \ref{fig:nanoscopic}a-b, vertical bars). The threshold correction is still larger in the mesoscopic regime as shown in Fig. \ref{fig:mesoscopic}a-b (vertical bars), 
leading to $\lambda^{(1)}_{a,th}=0.0627756$, in good agreement with the numerical threshold, $\lambda_{a,th}\approx 0.0647$; the correction $\delta^{(1)}$ is as large as $\sim 20\%$ of the class-A-like threshold ($\lambda^{(0)}_{a,th}=0.0536842$). 

\section{Dynamics of the second-order coherence}
\label{Correlations}

In the previous Section we have pointed out that, overall, the main difference between class-B solutions and class-A-like ones ---in which cavity-induced dissipation in the atomic variables as well as two-atom contributions in the light-matter interaction terms are neglected--- appears in the photon fluctuations. Remarkably, in the three regimes that we have numerically simulated---nanoscopic, mesoscopic and macroscopic--, $g^{(2)}(0)$ in the class-B model is larger than its class-A-like counterpart; above laser threshold, such a difference clearly means superpoissonian character of class-B light sources, specially in the nanoscopic and mesoscopic regimes. In this Section we will investigate the physical origin of such noise-excess by computing the time-lag dependence of the second order coherence, $g^{(2)}(\tau)$, from the dynamical equations [Eqs. \eqref{eq:modela}-\eqref{eq:modelP}]. 

\subsection{Time evolution of the second order coherence}

Let us consider the time-evolution operator $\hat U_{t,t_0}$ that evolves the density matrix from $t_0$ to $t$, 
\begin{equation}
\hat \rho(t)= \hat U_{t,t_0}[\hat \rho(t_0)].
\end{equation}
With the usual definition of the second-order correlation function  \cite{PhysRev.130.2529} $G(\tau,0)=\langle \hat a^{\dagger}_0 \hat a^{\dagger}_\tau \hat a_\tau \hat a_0 \rangle$, where $\hat a_\tau$ ($\hat a^{\dagger}_\tau$) is the annihilation (creator) operator in the Heisenberg representation at time $\tau$, it is possible to show that: 
\begin{equation}
G(\tau,0)=\mathrm{Tr} \left[ \hat a^{\dagger}  \hat a \hat U_{\tau,0}\left[ \hat a \hat \rho(0) \hat a^{\dagger}\right] \right].
\end{equation}
Defining a new operator $\hat \theta(\tau,0)=\hat U_{\tau,0} [\hat a \hat \rho(0) \hat a^{\dagger}]$, and setting $\hat \rho(0)=\hat \rho_{ss}$, where $\hat \rho_{ss}$ is the steady-state density matrix, it yields
\begin{equation}
\hat \theta(\tau,0)=\hat U_{\tau,0} \left[ \sum_{n',n''} \sqrt{n'n''} \rho_{ss,n',n''} |n''-1\rangle \langle n'-1|  \right],
\label{eq:theta}
\end{equation}
which represents the evolution of the density matrix once a photon has been emitted at $t=0$. We finally obtain: 
\begin{eqnarray}
g^{(2)}(\tau,0)&=& \frac{\mathrm{Tr} \left[ \hat a^{\dagger}  \hat a \hat U_{\tau,0}\left[ \hat a \hat \rho(0) \hat a^{\dagger}\right] \right]}{\mathrm{Tr} \left[ \hat a  \hat \rho(0)  \hat a^{\dagger}\right]  \mathrm{Tr} \left[ \hat a  \hat \rho(\tau) \hat a^{\dagger} \right]} \\
&=& \frac{\sum_n n \theta_{nn}(\tau,0)}{\mathrm{Tr} \left[ \hat a  \hat \rho(0)  \hat a^{\dagger} \right]\mathrm{Tr} \left[ \hat a  \hat \rho(\tau) \hat a^{\dagger} \right]}.
\end{eqnarray}

\subsection{Numerical results}

The time evolution of $\rho^a_n$ and $P_n$ governed by Eqs. \eqref{eq:modela}-\eqref{eq:modelP} gives a prescription to numerically solve the matrix elements of $\hat \theta(\tau,0)$ (Eq. \eqref{eq:theta}). First of all, $\rho^a_{ss,n}$ and $P_{ss,n}$ are obtained by running the Runge-Kutta algorithm for long simulation times,  $\rho^a_{ss,n}=\lim_{T\rightarrow \infty} \rho^a_n(T)$ and $P_{ss,n}=\lim_{T\rightarrow \infty} P_n(T)$. The Runge-Kutta algorithm is subsequently used to obtain the matrix elements $\theta_{nn}(\tau,0)=\langle n| \hat \theta(\tau,0) |n \rangle$ (Eq. \eqref{eq:theta}) by setting new initial conditions $\rho^a_n(0)\rightarrow (n+1) \rho^a_{ss,n}$ and $P_n(0)\rightarrow  (n+1) P_{ss,n}$ in Eqs. \eqref{eq:modela}-\eqref{eq:modelP}, in order to obtain, respectively,  $\theta^a_{nn}(\tau,0)$ and $\theta_{nn}(\tau,0)$, at a final integration time $T=\tau$.

\begin{figure}[t!]
\centering
\includegraphics[trim=0.cm 2cm 0cm 2.5cm,clip=true,scale=0.5,angle=0,origin=c]{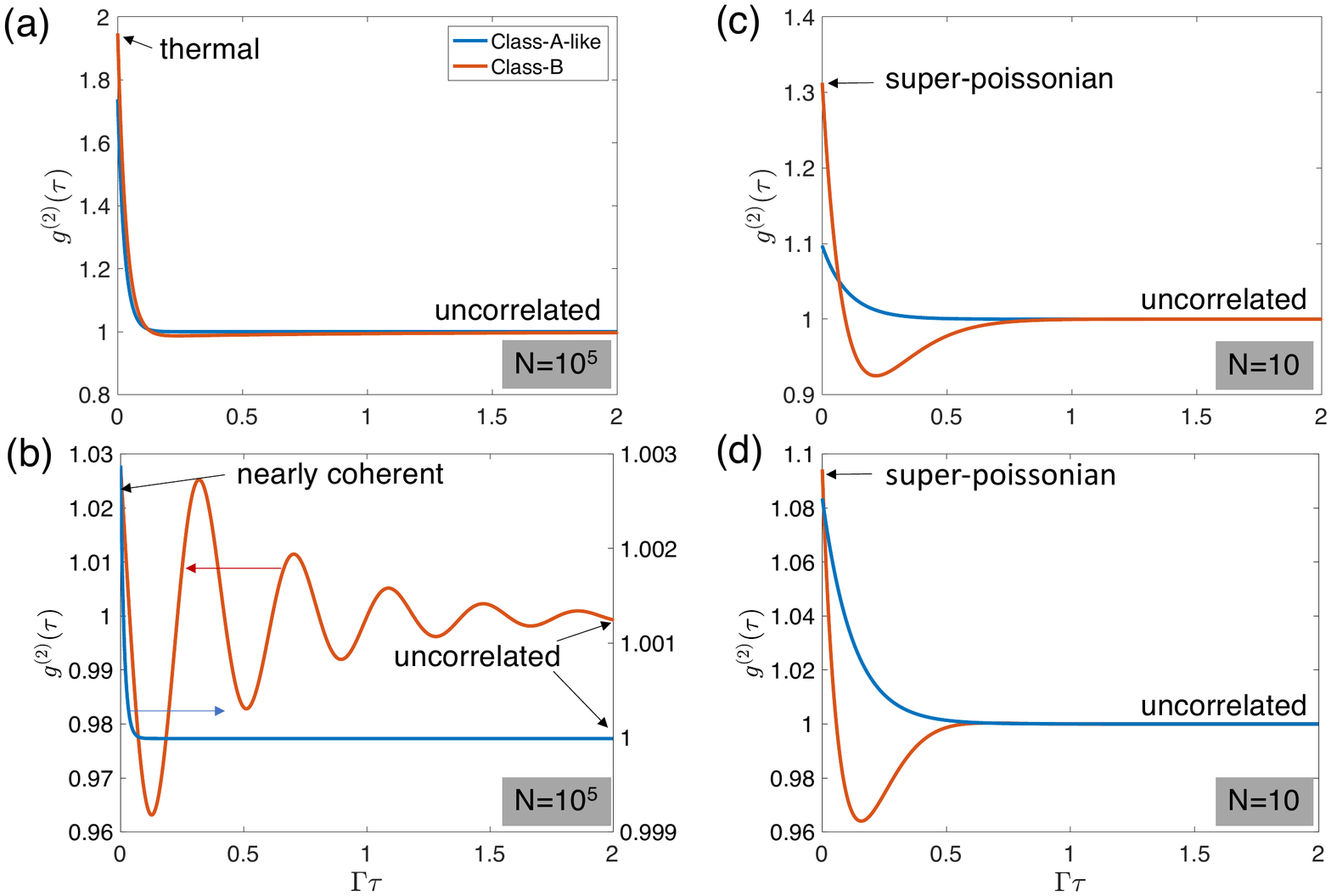}
\caption{Time-lag dependence of the second order coherence. (a)-(b) The mesoscopic regime of Fig. \ref{fig:mesoscopic}: (a) below threshold ($\lambda_a=0.039$), and (b) above threshold ($\lambda_a=0.15$). (c)-(d) The nanoscopic regime of Fig. \ref{fig:nanoscopic}: (c) below threshold ($\lambda_a=0.297$), and (d) above threshold ($\lambda_a=3$). Blue lines: class-A-like model; red lines: class-B model. Relaxation oscillations are clearly observed in the class-B model, and they are always absent in the class-A-like model. These are strongly damped in the nanoscopic regime.}
\label{fig:g2tau}
\end{figure}

Here we compute $g^{(2)}(\tau)$ for two of the several regimes investigated in Sec. \ref{Class-B numerics}: the mesoscopic and the nanoscopic ones (Figs. \ref{fig:mesoscopic} and \ref{fig:nanoscopic}, respectively). The $g^{(2)}(\tau)$ are computed using: i) the class-A-like model, and ii) the class-B model, and subsequently compared. The class-A-like solution is obtained numerically integrating Eq. \eqref{eq:modelP} with the adiabatically eliminated atomic reduced density matrix, resulting from neglecting both cavity-induced dissipation (${\cal  G}$-terms) and two-atom contributions (${\cal  F}$-terms) in Eq. \eqref{eq:modela}: 
\begin{equation}
\rho^a_n= \frac{\lambda_a P_n}{ \lambda_a+\Gamma+\frac{2g^2}{\gamma_h}(n+1) }.
\end{equation}

The results for the mesoscopic regime are shown in Fig.\ref{fig:g2tau}a-b. Below laser threshold (Fig.\ref{fig:g2tau}a) there is a monotonic decrease of the class-A-like $g^{(2)}(\tau)$ towards the uncorrelated $g^{(2)}(\tau)\rightarrow 1$ limit for large $\tau$. As long as the class-B-solution is concerned, $g^{(2)}(\tau)$ falls slightly below $g^{(2)}=1$ and slowly increases towards the uncorrelated limit $g^{(2)}(\tau)\rightarrow 1$ for large $\tau$. In spite of this slight difference, the two solutions are close to each other. 
However, there is a remarkable difference between the two models above threshold (Fig.\ref{fig:g2tau}b): while the class-A-like model predicts a monotonic decrease of $g^{(2)}(\tau)$, the class-B model solution is oscillatory. We interpret such damped oscillations as the autocorrelation of relaxation oscillations in a class-B laser. In a quantum trajectory picture ---also in intensity-time series of classical representations---, the relaxation oscillations are transients, permanently triggered by spontaneous emission noise. The noise strength impacts not only $g^{(2)}(0)$ but also the damping time of the oscillations. We observe that, as a general trend, the smaller $\beta$-factors, the longer decay time of relaxation oscillations in the autocorrelation trace. 
This behavior of the relaxation time at the threshold can indeed be interpreted as a slowing down ---which become critical in the thermodynamic limit $\beta\to0$--- of the amplitude mode, as observed also for class-A lasers \cite{Takemura_2021}.
This trend can be connected with the spectral properties of the Liouvillian \cite{Minganti_2018} ruling the open system dynamics. 

\begin{figure}[t!]
\centering
\includegraphics[trim=4cm 3cm 0cm 2.5cm,clip=true,scale=0.7,angle=0,origin=c]{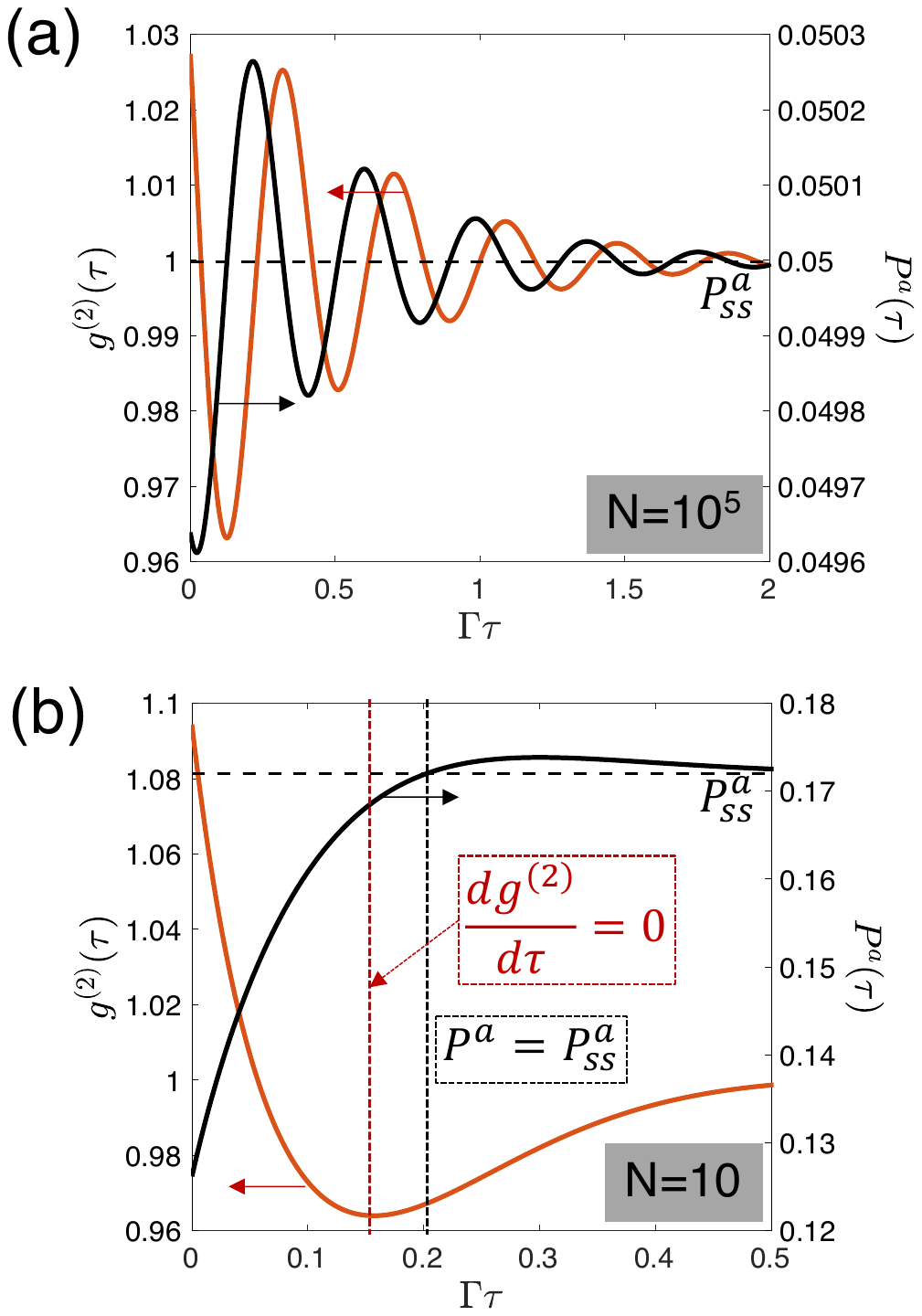}
\caption{Relaxation oscillations in a class-B laser: from classical to few photon regimes. Left axis: time-lag dependent second order coherence. Right axis: time-dependent occupation probability of the upper laser state $|a \rangle$ once a photon has been emitted at $t=0$. (a) Mesoscopic regime (paramaters of Fig. \ref{fig:mesoscopic}), $\lambda_a=0.2$. The time shift between the signals assets the classical relaxation oscillation character of the autocorrelation trace: the occupation probability crosses the steady state value (dashed line) when the $g^{(2)}$ reaches an extremum. (b) Nanoscopic, few photon regime (paramaters of Fig. \ref{fig:nanoscopic}), $\lambda_a=3$. Spontaneous emission cannot be neglected in this few-photon regime; the extremum of photon probability (red vertical line) does not correspond to the steady state value of the occupation probability (black vertical line).}
\label{fig:g2tau_theta}
\end{figure}

In Fig. \ref{fig:g2tau}c-d we display the results of the nanoscopic regime: $\beta\sim 0.6-0.8$ and $N=10$. Clearly, in this case there is no laser threshold in the sense of phase transition. Let us recall that the laser threshold at about $\lambda_a\sim 1.5$ is a crossover from $P_0> P_1$ to $P_0<P_1$. Yet, as it is well-known in the so-called \lq\lq thresholdless regime'' of micro and nanolasers, the emission properties do not substantially differ below and above threshold ---if any. This is also the case here: although there is a monotonic decrease of $g^{(2)}(0)$ for increasing pump, there is no sharp decrease from $g^{(2)}(0)=2$ to  $g^{(2)}(0)=1$ as it is clearly observed in the macroscopic regime (Fig. \ref{fig:single-thermo}d). The $g^{(2)}(\tau)$ is also similar when comparing below and above-threshold (Fig. \ref{fig:g2tau}c and d, respectively). Although the class-B model predicts superpoissonian light, there are no clear relaxation oscillations; instead, a downwards bump at $\Gamma  \tau\sim 0.2$ followed by an increase towards the $g^{(2)}=1$ limit suggests that relaxation oscillations are overdamped in this regime. We observe that overdamping of relaxation oscillations is characteristic of large $\beta$-factors, and it is particularly strong for reduced atomic numbers as in this case. 

Relaxation oscillations is a classical concept, that we are hereby extending into the quantum domain. It is well known that, in a class-B-laser semiclassical model, the laser intensity and the population inversion both exhibit damped oscillations. The energy is actually exchanged between photon and atom populations before reaching equilibrium. These oscillations are time-lagged: the  laser intensity reaches an extremum each time the population inversion crosses its steady state value. This can be easily deduced from a standard rate equation for photons: $\dot s=\beta_0 \gamma_\parallel {\cal N}(s+1)-2\kappa s$, where $s$ is the photon number, ${\cal N}$ is the population inversion and $\gamma_\parallel=\gamma+4g^2/\gamma_h$ is the total radiative recombination time. Neglecting spontaneous emission ($s\gg 1$), the steady state solution reads ${\cal N}_{th}=2\kappa/\beta_0 \gamma_\parallel=\kappa\gamma_h/2g^2$. Within a transient relaxation to the steady state, the intensity dynamics is expected to reach an extremum ($\dot s=0$) when the population inversion crosses its steady state value ${\cal N}={\cal N}_{th}$. In Fig. \ref{fig:g2tau_theta} we depict both  $g^{(2)}(\tau)$ and $P^a(\tau)\equiv \mathrm{Tr}[\hat \theta^a(\tau,0)]/\langle n \rangle$, the latter accounting for the time evolution of the occupation probability of upper laser lever. In the mesoscopic regime (Fig.  \ref{fig:g2tau_theta}a) the photon number is large enough ($s\sim 150$) such that the spontaneous emission can be neglected and therefore we can expect relaxation oscillations to exhibit classical features. Indeed, the time-lag between both correlation traces is in agreement with the classical picture. In the nanoscopic regime, on the other hand, the time-lag between correlation traces does not agree with the classical picture (Fig. \ref{fig:g2tau_theta}b, vertical dotted lines). Systematic studies on the relaxation oscillation properties varying the laser parameters, including large vs. small $\beta$-factors, and large vs. small $N$, is out of the scope of the present work and is left for future investigations.

As a conclusion of this Section, we have shown that the superpoissonian light in this class-B model is originated from photon-number-noise excess due to damped relaxation oscillations. This seems in good agreement with previous experimental studies, in which an interferometric Hanboury Brown-Twiss set-up enables to discriminate phase and intensity fluctuations \cite{PhysRevLett.110.163603}. The authors conclude that, in nanolasers, the excess of $g^{(2)}(0)$ above the Poissonian limit, which tends very slowly towards coherence upon pump power increase ---in agreement with Fig. \ref{fig:mesoscopic}b---, is due to intensity fluctuations, yet the phase coherence is preserved. First-order correlations, related to field coherence and hence to phase noise, can also be computed with our approach. This requires working out non diagonal elements of the density matrix, which would be a natural extension of the present work.

\section{Conclusions}
\label{Conclusions}
In this work we have extended the quantum density matrix approach for a laser developed by M. Scully and co-workers in the 80s-90s, in which atomic variables can be adiabatically eliminated (class-A), to the regime in which adiabatic elimination is no longer possible because the atomic decay rates are smaller than the cavity dissipation rate (class-B). Class-B laser regimes encompass the great majority of solid state lasers, from ion-doped crystals to semiconductors, and all the way from the nanoscale (few photons, few "atoms"), to the thermodynamic regime (very large photon and atom numbers). We show that the following two approximations, that greatly simplify the model and enable analytical treatment for class-A regimes, no longer apply for class-B: i) nelglecting the cavity-induced dissipation in the atomic reduced density matrix equations, and ii) neglecting two-atom contributions to the gain. As a result, instead of a single photonic equation of motion for class-A lasers, our class-B model deals with two coupled dynamical equations, one for the photonic and one for the atomic reduced density matrix elements, reminiscent of the semiclassical two-variable models for the electromagnetic field and population inversion.
Our model is a nonlinear dynamical system of equations extended in a pseudo-space dimension given by the photon number, and can be numerically solved using standard numerical integration methods such as the Fourth-order Runge-Kutta algorithm. In addition, approximate analytical expressions can be derived for the laser threshold. Interestingly, this approach allows us to define the laser threshold in a universal way: the probability for one intracavity photon equals that of the quantum vacuum, $P_0=P_1$.  

This quantum class-B model enables to describe the full statistics of the laser transition in very different operation regimes, and accounts for fundamental physical phenomena, from photon antibunching in single-atom cavities to the thermal-to-coherent-light phase transition as both the photon and atomic numbers are very large (thermodynamic limit). In addition, also nano and mesoscale phenomena are well captured by the class-B quantum model, ranging from the thresholdless regime in the former, to the typical "S" input/output logaritmic laser curves in micro and nanolasers. This class-B model predicts superpoissonian light, $g^{(2)}(\tau=0)>1$, even far above the laser threshold. We have related this super-poissonian light to amplitude-noise excess due to relaxation oscillations which are known to be ubiquitous in class-B lasers, in agreement with Ref.  \cite{PhysRevLett.110.163603}. Unlike the class-A model, we generically observe relaxation oscillations as damped oscillatory behavior of $g^{(2)}(\tau)$ in the class-B model; $g^{(2)}(\tau)$  can be readily obtained from the time evolution of an operator $\hat \theta(\tau,0)$ that accounts for the probability of photon emission at $t=\tau$ provided a photon has been emitted at $t=0$. We find that the oscillatory $g^{(2)}(\tau)$ exhibit the classical features of relaxation oscillations in the large photon number regime ($n\gg n_{sat}$) where the spontaneous emission can be neglected, in the form of periodic energy exchange between photons and atoms. However, oscillations become overdamped and the time-lag between photons and population inversion is no longer predicted by the semiclassical dynamical rules in few photon laser regimes.

This work also paves the way to future intriguing research directions. 
Among them we mention the characterization of atom-atom quantum correlations in class-B lasers. As we discussed in this work the non-factorizable nature of the system density matrix is a key ingredient in triggering the lasing mechanism, however it is not clear if quantum or classical correlations play this crucial role and if the lasing transition is connected to some sharp change in the entanglement properties of the atoms.  
Furthermore, a systematic connection of transition to lasing in class-B systems and dissipative phase transitions \cite{Minganti_2018,Minganti_2021} represents an interesting direction to be further explored. Here we expect that the behaviour of the Liouvillian gap signaling the emergence of a dissipative phase transition ---lasing, in this case--- can be related to the decay time of the first-order coherence (as done in Ref. \cite{Takemura_2021} for a class-A laser), whose calculation requires to compute off-diagonal elements of the photonic density matrix. Finally, this quantum density matrix approach proves a powerful theoretical framework to investigate few photon nonlinear and quantum phenomena, such as photon entanglement, in coherently coupled nanolaser arrays.

\section*{acknowledgements}
We acknowledge  N. Bartolo , F. Vicentini and F. Minganti for useful discussions. This work was supported by ANR via the projects UNIQ (ANR-16-CE24-0029), NOMOS (ANR-18-CE24-0026), TRIANGLE (ANR-20-CE47-0011), the FET FLAGSHIP Project PhoQuS (grant agreement ID: 820392) and  the European Union in the form of Marie Sk\l odowska-Curie Action grant MSCA-841351.

\section*{conflict of interest}
The authors declare no conflicts of interest.

\bibliography{sample.bib}

\begin{thebibliography}{24}
\providecommand{\natexlab}[1]{#1}
\providecommand{\url}[1]{\texttt{#1}}
\providecommand{\urlprefix}{}

\bibitem[{Scully and Zubairy(1997)Scully, Marlan O and Zubairy, M
  Suhail}]{scully1997quantum}
Scully MO, Zubairy MS.
\newblock Quantum theory of the laser--density operator approach.
\newblock Quantum Optics (Cambridge University Press, 1997) pp 1997;p.
  327--361.

\bibitem[{Takemura et~al.(2021)Naotomo Takemura and Masato Takiguchi and Masaya
  Notomi}]{Takemura_2021}
Takemura N, Takiguchi M, Notomi M.
\newblock Low- and high-$\beta$ lasers in the class-A limit: photon statistics,
  linewidth, and the laser-phase transition analogy.
\newblock J Opt Soc Am B 2021 Mar;38(3):699--710.
\newblock
  \urlprefix\url{http://opg.optica.org/josab/abstract.cfm?URI=josab-38-3-699}.

\bibitem[{Arecchi and Harrison(2012)Arecchi, F Tito and Harrison, Robert
  G}]{arecchi2012instabilities}
Arecchi FT, Harrison RG.
\newblock Instabilities and chaos in quantum optics, vol.~34.
\newblock Springer Science \& Business Media; 2012.

\bibitem[{Narducci and Abraham(1988)Narducci, Lorenzo M and Abraham, Neal
  B}]{narducci1988laser}
Narducci LM, Abraham NB.
\newblock Laser physics and laser instabilities.
\newblock World Scientific; 1988.

\bibitem[{Rice and Carmichael(1994)Rice, Perry R. and Carmichael, H.
  J.}]{PhysRevA.50.4318}
Rice PR, Carmichael HJ.
\newblock Photon statistics of a cavity-QED laser: A comment on the
  laser--phase-transition analogy.
\newblock Phys Rev A 1994 Nov;50:4318--4329.
\newblock \urlprefix\url{https://link.aps.org/doi/10.1103/PhysRevA.50.4318}.

\bibitem[{Yamamoto et~al.(1991)Yamamoto, Y. and Machida, S. and Bj\"ork,
  G.}]{PhysRevA.44.657}
Yamamoto Y, Machida S, Bj\"ork G.
\newblock Microcavity semiconductor laser with enhanced spontaneous emission.
\newblock Phys Rev A 1991 Jul;44:657--668.
\newblock \urlprefix\url{https://link.aps.org/doi/10.1103/PhysRevA.44.657}.

\bibitem[{Protsenko et~al.(1999)Protsenko, I. and Domokos, P. and
  Lef\`evre-Seguin, V. and Hare, J. and Raimond, J. M. and Davidovich,
  L.}]{PhysRevA.59.1667}
Protsenko I, Domokos P, Lef\`evre-Seguin V, Hare J, Raimond JM, Davidovich L.
\newblock Quantum theory of a thresholdless laser.
\newblock Phys Rev A 1999 Feb;59:1667--1682.
\newblock \urlprefix\url{https://link.aps.org/doi/10.1103/PhysRevA.59.1667}.

\bibitem[{Leymann et~al.(2013)Leymann, H. A. M. and Hopfmann, C. and Albert, F.
  and Foerster, A. and Khanbekyan, M. and Schneider, C. and H\"ofling, S. and
  Forchel, A. and Kamp, M. and Wiersig, J. and Reitzenstein,
  S.}]{PhysRevA.87.053819}
Leymann HAM, Hopfmann C, Albert F, Foerster A, Khanbekyan M, Schneider C,
  et~al.
\newblock Intensity fluctuations in bimodal micropillar lasers enhanced by
  quantum-dot gain competition.
\newblock Phys Rev A 2013 May;87:053819.
\newblock \urlprefix\url{https://link.aps.org/doi/10.1103/PhysRevA.87.053819}.

\bibitem[{Moelbjerg et~al.(2013)Moelbjerg, Anders and Kaer, Per and Lorke,
  Michael and Tromborg, Bjarne and Mørk, Jesper}]{6603264}
Moelbjerg A, Kaer P, Lorke M, Tromborg B, Mørk J.
\newblock Dynamical Properties of Nanolasers Based on Few Discrete Emitters.
\newblock IEEE Journal of Quantum Electronics 2013;49(11):945--954.

\bibitem[{Gies et~al.(2007)Gies, Christopher and Wiersig, Jan and Lorke,
  Michael and Jahnke, Frank}]{PhysRevA.75.013803}
Gies C, Wiersig J, Lorke M, Jahnke F.
\newblock Semiconductor model for quantum-dot-based microcavity lasers.
\newblock Phys Rev A 2007 Jan;75:013803.
\newblock \urlprefix\url{https://link.aps.org/doi/10.1103/PhysRevA.75.013803}.

\bibitem[{Wiersig et~al.(2009)Wiersig, Jan and Gies, Christopher and Jahnke,
  Frank and Aßmann, Marc and Berstermann, Thorsten and Bayer, Manfred and
  Kistner, Caroline and Reitzenstein, S. and Schneider, Christian and Höfling,
  Sven and Forchel, Alfred and Kruse, Carsten and Kalden, Joachim and Hommel,
  D.}]{Wiersig_2009}
Wiersig J, Gies C, Jahnke F, Aßmann M, Berstermann T, Bayer M, et~al.
\newblock Direct observation of correlations between individual photon emission
  events of a microcavity laser.
\newblock Nature 2009 08;460:245--9.

\bibitem[{Kira et~al.(1999)M. Kira and F. Jahnke and W. Hoyer and S.W.
  Koch}]{KIRA1999189}
Kira M, Jahnke F, Hoyer W, Koch SW.
\newblock Quantum theory of spontaneous emission and coherent effects in
  semiconductor microstructures.
\newblock Progress in Quantum Electronics 1999;23(6):189--279.
\newblock
  \urlprefix\url{https://www.sciencedirect.com/science/article/pii/S0079672799000087}.

\bibitem[{Chow et~al.(2014)Chow, Weng W and Jahnke, Frank and Gies,
  Christopher}]{chow2014emission}
Chow WW, Jahnke F, Gies C.
\newblock Emission properties of nanolasers during the transition to lasing.
\newblock Light: Science \& Applications 2014;3(8):e201--e201.

\bibitem[{{Kreinberg} et~al.(2017){Kreinberg}, S{\"o}ren and {Chow}, Weng W.
  and {Wolters}, Janik and {Schneider}, Christian and {Gies}, Christopher and
  {Jahnke}, Frank and {H{\"o}fling}, Sven and {Kamp}, Martin and
  {Reitzenstein}, Stephan}]{2017LSA.....6E7030K}
{Kreinberg} S, {Chow} WW, {Wolters} J, {Schneider} C, {Gies} C, {Jahnke} F,
  et~al.
\newblock {Emission from quantum-dot high-{\ensuremath{\beta}} microcavities:
  transition from spontaneous emission to lasing and the effects of
  superradiant emitter coupling}.
\newblock Light: Science \& Applications 2017 Feb;6(8):e17030.

\bibitem[{Lorke et~al.(2013)Lorke, M. and Suhr, T. and Gregersen, N. and
  M\o{}rk, J.}]{PhysRevB.87.205310}
Lorke M, Suhr T, Gregersen N, M\o{}rk J.
\newblock Theory of nanolaser devices: Rate equation analysis versus
  microscopic theory.
\newblock Phys Rev B 2013 May;87:205310.
\newblock \urlprefix\url{https://link.aps.org/doi/10.1103/PhysRevB.87.205310}.

\bibitem[{Carroll et~al.(2021)Carroll, Mark Anthony and D'Alessandro, Giampaolo
  and Lippi, Gian Luca and Oppo, Gian-Luca and Papoff,
  Francesco}]{PhysRevLett.126.063902}
Carroll MA, D'Alessandro G, Lippi GL, Oppo GL, Papoff F.
\newblock Thermal, Quantum Antibunching and Lasing Thresholds from Single
  Emitters to Macroscopic Devices.
\newblock Phys Rev Lett 2021 Feb;126:063902.
\newblock
  \urlprefix\url{https://link.aps.org/doi/10.1103/PhysRevLett.126.063902}.

\bibitem[{Minganti et~al.(2018)Minganti, Fabrizio and Biella, Alberto and
  Bartolo, Nicola and Ciuti, Cristiano}]{Minganti_2018}
Minganti F, Biella A, Bartolo N, Ciuti C.
\newblock Spectral theory of Liouvillians for dissipative phase transitions.
\newblock Phys Rev A 2018 Oct;98:042118.
\newblock \urlprefix\url{https://link.aps.org/doi/10.1103/PhysRevA.98.042118}.

\bibitem[{Minganti et~al.(2021)Minganti, Fabrizio and Arkhipov, Ievgen I. and
  Miranowicz, Adam and Nori, Franco}]{Minganti_2021}
Minganti F, Arkhipov II, Miranowicz A, Nori F.
\newblock Liouvillian spectral collapse in the Scully-Lamb laser model.
\newblock Phys Rev Research 2021 Dec;3:043197.
\newblock
  \urlprefix\url{https://link.aps.org/doi/10.1103/PhysRevResearch.3.043197}.

\bibitem[{Lebreton et~al.(2013)Lebreton, A. and Abram, I. and Braive, R. and
  Sagnes, I. and Robert-Philip, I. and Beveratos, A.}]{PhysRevLett.110.163603}
Lebreton A, Abram I, Braive R, Sagnes I, Robert-Philip I, Beveratos A.
\newblock Unequivocal Differentiation of Coherent and Chaotic Light through
  Interferometric Photon Correlation Measurements.
\newblock Phys Rev Lett 2013 Apr;110:163603.
\newblock
  \urlprefix\url{https://link.aps.org/doi/10.1103/PhysRevLett.110.163603}.

\bibitem[{Breuer and Petruccione(2002)Breuer, H. P. and Petruccione,
  F.}]{PetruccioneOQS}
Breuer HP, Petruccione F.
\newblock The theory of open quantum systems.
\newblock Great Clarendon Street: Oxford University Press; 2002.

\bibitem[{{Wang} et~al.(2015){Wang}, T. and {Puccioni}, G.~P. and {Lippi},
  G.~L.}]{2015NatSR...515858W}
{Wang} T, {Puccioni} GP, {Lippi} GL.
\newblock {Dynamical Buildup of Lasing in Mesoscale Devices}.
\newblock Scientific Reports 2015 Oct;5:15858.

\bibitem[{Marconi et~al.(2020)Marconi, Mathias and Raineri, Fabrice and
  Levenson, Ariel and Yacomotti, Alejandro M. and Javaloyes, Julien and Pan, Si
  H. and Amili, Abdelkrim El and Fainman, Yeshaiahu}]{PhysRevLett.124.213602}
Marconi M, Raineri F, Levenson A, Yacomotti AM, Javaloyes J, Pan SH, et~al.
\newblock Mesoscopic Limit Cycles in Coupled Nanolasers.
\newblock Phys Rev Lett 2020 May;124:213602.
\newblock
  \urlprefix\url{https://link.aps.org/doi/10.1103/PhysRevLett.124.213602}.

\bibitem[{Perea et~al.(2004)Perea, J. I. and Porras, D. and Tejedor,
  C.}]{PhysRevB.70.115304}
Perea JI, Porras D, Tejedor C.
\newblock Dynamics of the excitations of a quantum dot in a microcavity.
\newblock Phys Rev B 2004 Sep;70:115304.
\newblock \urlprefix\url{https://link.aps.org/doi/10.1103/PhysRevB.70.115304}.

\bibitem[{Glauber(1963)Glauber, Roy J.}]{PhysRev.130.2529}
Glauber RJ.
\newblock The Quantum Theory of Optical Coherence.
\newblock Phys Rev 1963 Jun;130:2529--2539.
\newblock \urlprefix\url{https://link.aps.org/doi/10.1103/PhysRev.130.2529}.

\end{thebibliography}

%
%
%

\end{document}